\documentclass[12pt,final,a4paper]{iopart}
\usepackage{iopams}
\usepackage{epsfig}
\usepackage{array}
\usepackage{longtable}
\usepackage{rotating}
\usepackage{subfigure}
\usepackage{amssymb}
\usepackage{mathrsfs}
\usepackage{calrsfs}
\usepackage{graphics}
\usepackage{txfonts}
\usepackage{pslatex}
\newcommand{\sqrtsnn}{\sqrt{s_{\mbox{\tiny{\it{NN}}}}}}
\newcommand{\sqrts}{\sqrt{s}}
\def\mean#1{\ensuremath{\left<#1\right>}}

\newcommand\qhat{{\mean{\hat{q}}}}
\newcommand{\jpsi}{J/\psi}

\newcommand{\ups}{\Upsilon}

\newcommand{\dNdeta}{dN_{ch}/d\eta|_{\eta=0}}
\newcommand{\dETdeta}{dE_{T}/d\eta|_{\eta=0}}
\newcommand{\ecrit}{\varepsilon_{\mbox{\tiny{\it crit}}}}
\newcommand{\ebj}{\varepsilon_{\mbox{\tiny{\it Bj}}}}
\newcommand{\Tcrit}{T_{\mbox{\tiny{\it crit}}}}
\newcommand{\Teff}{T_{\mbox{\tiny{\it eff}}}}
\newcommand{\geff}{g_{\mbox{\tiny{\it eff}}}}

\begin{document}

\setcounter{footnote}{1}

\topical[Quark-Gluon Matter]
{Quark-Gluon Matter}

\author{David~d'Enterria}
\address{CERN, PH-EP, CH - 1211 Geneva 23, Switzerland}
\ead{e-mail:david.d'enterria@cern.ch}

%X pages, X figures, invited overview paper (J. Phys. G: Nucl. Part. Phys.)

%%%%%%%%%%%%%%%%%%%%%%%%%%%%%%%%%%%%%%%%%%%%%%%%%%%%%%%%%%%%%%%%

\begin{abstract}
A concise review of the experimental and phenomenological progress in high-energy heavy-ion 
physics over the past few years is presented. Emphasis is put on measurements 
at BNL-RHIC and CERN-SPS which provide information on fundamental properties 
of QCD matter at extreme values of temperature, density and low-$x$. The new 
opportunities accessible at the LHC, which may help clarify some of the current open 
issues, are also outlined.
\end{abstract}

\submitto{{\it J. Phys.~G: Nucl. Part. Phys.}}

%\maketitle

%%%%%%%%%%%%%%%%%%%%%%%%%%%%%%%%%%%%%%%%%%%%%%%%%%%%%%%%%%%%%%%%

\section*{Introduction}
\label{sec:intro}

The study of the fundamental theory of the strong interaction -- Quantum Chromo Dynamics (QCD) --
in extreme conditions of temperature, density and small parton momentum fraction (low-$x$) 
has attracted an increasing experimental and theoretical interest during the last 20 years. 
Indeed, QCD is not only a quantum field theory with an extremely rich dynamical content 
(asymptotic freedom, infrared slavery, (approximate) chiral symmetry, non trivial vacuum topology, 
strong CP violation problem, $U_A(1)$ axial-vector anomaly, ...) but also the only sector of the 
Standard Model whose full {\it collective} behaviour -- phase diagram, phase transitions, thermalization
of fundamental fields -- is accessible to scrutiny in the laboratory. The study of the many-body dynamics 
of high-density QCD covers a vast range of fundamental physics problems (Fig.~\ref{fig:QCD_facets}):

\begin{figure}[htbp]
\centering
\hskip 0.8cm
\includegraphics[width=6.5cm]{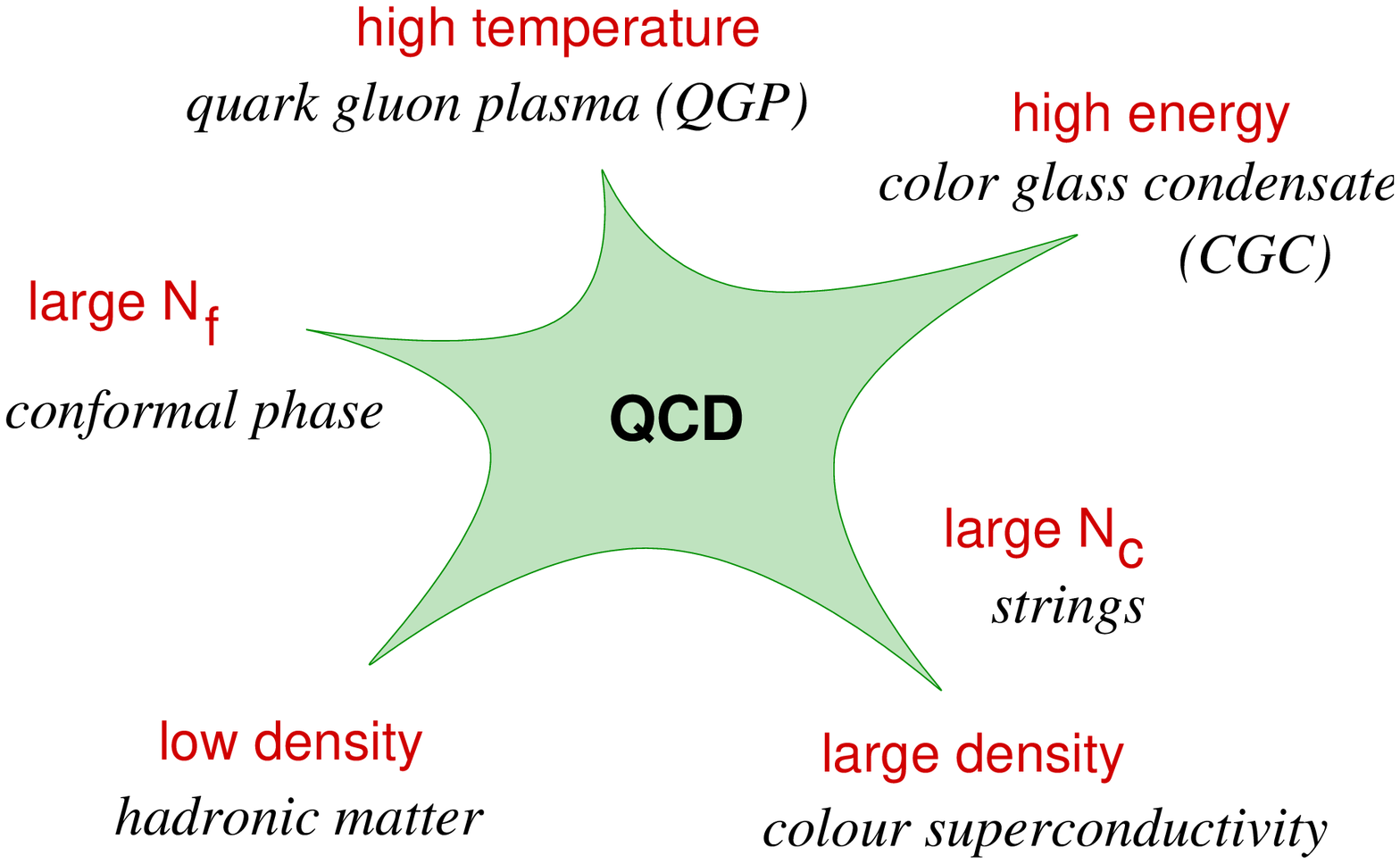}
\includegraphics[width=8.2cm,height=6.cm]{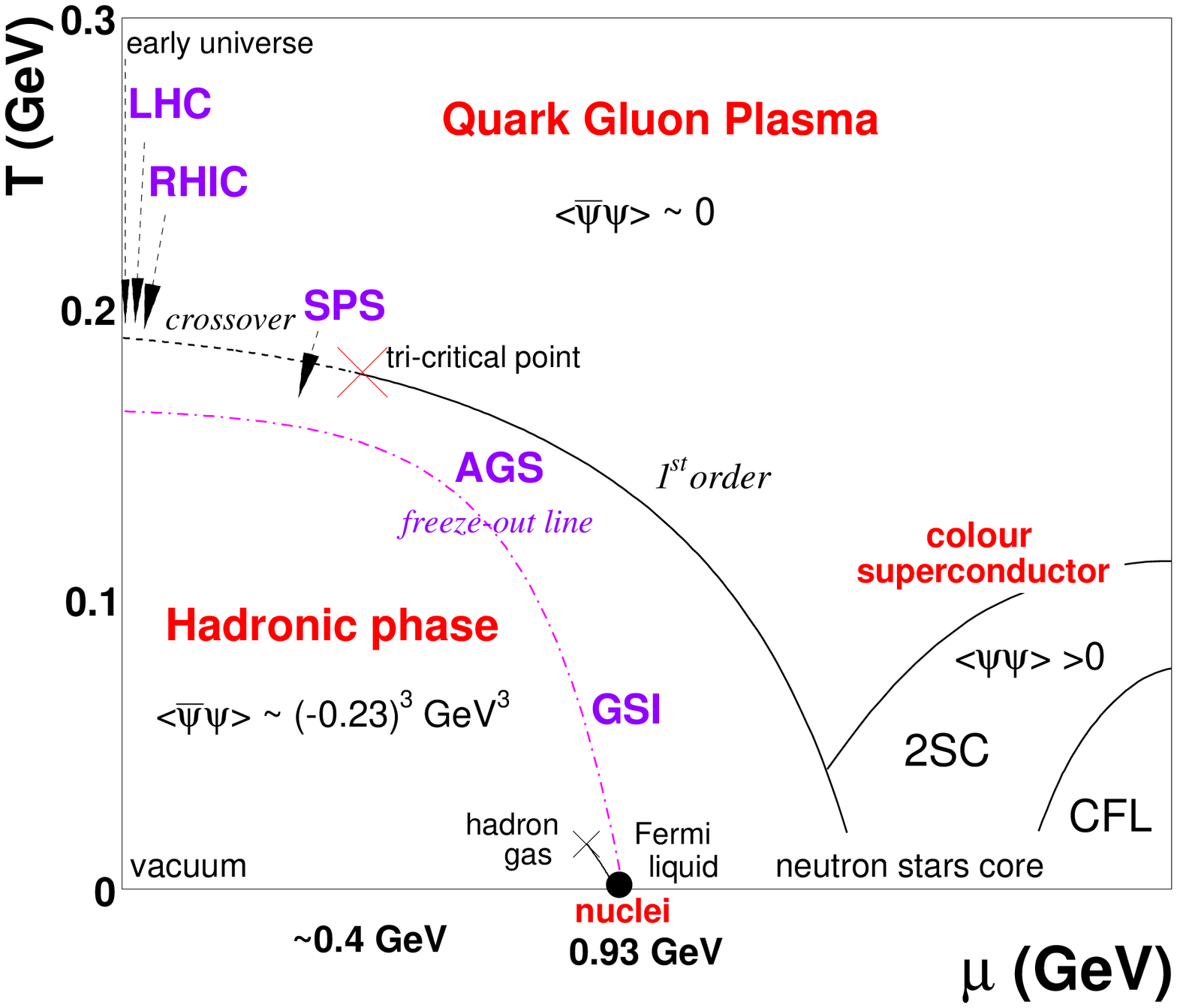}
%\vskip -0.7cm
\caption{Left: The multiple facets of QCD (adapted from~\protect\cite{schaefer05}). 
Right: QCD phase diagram in the temperature vs. baryochemical potential ($T,\mu_B$) plane.
The arrows indicate the expected crossing through the deconfinement transition during the 
expansion phase in heavy-ion collisions at different accelerators. The dashed freeze-out 
curve indicates where hadrochemical equilibrium is attained in the latest stage of 
the collision~\protect\cite{freezeout06}.}
\label{fig:QCD_facets}
\end{figure} 

\begin{itemize}
\item {\bf Deconfinement and chiral symmetry restoration}:
Lattice QCD calculations~\cite{latt} predict a new form of matter at energy densities (well) 
above $\ecrit \approx$ 1 GeV/fm$^3$ consisting of an extended volume of deconfined and 
%$\ecrit=(6\pm 2)\Tcrit^4\approx$ 1 GeV/fm$^3$ consisting of an extended volume of deconfined and 
bare-mass quarks and gluons: the Quark Gluon Plasma (QGP)~\cite{shuryak77}. The scrutiny of this new 
state of matter -- equation-of-state (EoS), order of the phase transition, transport properties, etc. -- promises to 
shed light on basic aspects of the strong interaction such as the nature of confinement, 
the mechanism of mass generation (chiral symmetry breaking, structure of the QCD vacuum) and hadronization, 
which still evade a thorough theoretical description~\cite{millenium_prizes} due to their highly non-perturbative 
nature.

\item {\bf Early universe cosmology}: The quark-hadron phase transition took place some 10 $\mu$s 
after the Big-Bang and is believed to have been the most important event in the Universe %taking place 
between the electro-weak (or SUSY) transition ($\tau\sim 10^{-10}$ s) and nucleosynthesis 
($\tau\sim$ 200 s). Depending on the order of the transition\footnote{The order itself is not exactly known: 
the transition, which is 1$^{st}$-order in pure SU(3) gluodynamics and of a fast cross-over type
for $N_f$ = 2+1 quarks~\cite{latt}, is still sensitive to lattice extrapolations to the continuum limit.}, 
several cosmological implications have been postulated~\cite{Schwarz:2003du} such as the formation 
of strangelets and cold dark-matter (WIMP) clumps, or baryon fluctuations leading to inhomogeneous 
nucleosynthesis.

\item {\bf Parton structure and evolution at small-$x$}: HERA data~\cite{hera_lowx} indicate that when 
probed at high energies, hadrons consist of a very dense system of gluons with small (Bjorken)  momentum 
$x=p_{\mbox{\tiny{\it parton}}}/p_{\mbox{\tiny{\it hadron}}}$. 
At low $x$, the probability to emit an extra gluon is large, proportional to $\alpha_s\ln(1/x)$, 
and $gg$ fusion processes will play an increasing role in the parton evolution in the hadronic wavefunctions. 
At high virtualities $Q^2$ and moderately low $x$, such evolution %with $Q^2$ (or $\ln(1/x)$) 
is described by linear DGLAP~\cite{dglap} or BFKL~\cite{bfkl} equations, suitable for a dilute parton
regime. At $x\lesssim 10^{-2}$ and below an energy-dependent ``saturation momentum'' $Q_s$, 
hadrons are however more appropriately %$Q^2_s \approx\alpha_s\,xG(x,Q^2)/(\pi\,R^2)$, such a configuration 
described as dense, saturated parton systems in the context of the ``Colour Glass Condensate'' 
(CGC)~\cite{iancu03} effective theory with the corresponding non-linear BK/JIMWLK~\cite{bk,jimwlk} 
evolution equations. Since the growth of the gluon density depends on the transverse size 
of the hadron, saturation effects are expected to set in earlier for ultrarelativistic heavy nuclei 
%(for which $Q_s^2\propto A^{1/3}$, where $A$ is the number of nucleons) than for free nucleons.
(for which $Q_s^2\propto A^{1/3}$, with $A$ the number of nucleons) than for free nucleons.

\item  {\bf Gauge/String duality}: Theoretical applications of the Anti-de-Sitter/Conformal-Field-Theory 
(AdS/CFT) correspondence provide results in strongly coupled (i.e.\ large 't~Hooft coupling
$\lambda = g^2\,N_c\gg 1$) SU($N_c$) gauge theories in terms of a weakly-coupled dual gravity 
theory~\cite{ads_cft}. Recent applications of this formalism for QCD-like (${\cal N}=4$ super Yang-Mills) %(SYM) 
theories have led to the determination of  transport properties of experimental relevance -- such as the QGP 
viscosity~\cite{kovtun04}, the ``jet quenching'' parameter $\qhat$~\cite{wiedem06}, 
or the heavy-quark diffusion coefficient~\cite{heavyQ_adscft} -- %or different medium spectral functions
from black hole thermodynamics calculations. Such results provide valuable insights on {\it dynamical} 
properties of strongly-coupled QCD that cannot be directly treated by perturbative or lattice methods, 
and open new phenomenological and experimental leads.

\item  {\bf Compact object astrophysics}: At high baryon densities and not too high temperatures, the attractive 
force between (colour antisymmetric) quarks can lead to the formation of bound $\mean{qq}$ %condensates of 
Cooper pairs. Cold dense matter is thus expected to behave as a colour super-conductor %, a Fermi liquid 
%At ultra-high density, matter is expected to form a degenerate Fermi gas of
%quarks in which there is a condensate of Cooper pairs of quarks near the Fermisurface.
with a non trivial quark pairing structure due to the combination of the various 
quantum numbers involved (spin, colour, flavour)~\cite{colour_supercond}. 
This regime, currently beyond the direct reach of accelerator-based research (except indirectly
in the region of baryon densities around the QCD tri-critical point, Fig.~\ref{fig:QCD_facets} right), 
may be realised in the core of compact (neutron, hybrid or other exotic) stars and, thus, open to study
through astronomical observation.
%and in the violent events associated with collapse of massive stars or collisions of neutron stars.
%The hyperdense conditions prevailing in the core of compact stars (white dwarfs, neutron or exotic stars) are such 
%In the range of baryon densities experimentally accessible in heavy-ion collisions, lattice QCD calculations
%indicate the existence of a tricritical point in the temperature versus baryochemical potential plane~\cite{}. 
\end{itemize}

\begin{figure}[!htp]
\centering
\hskip -0.5cm
\includegraphics[width=7.8cm,height=6cm]{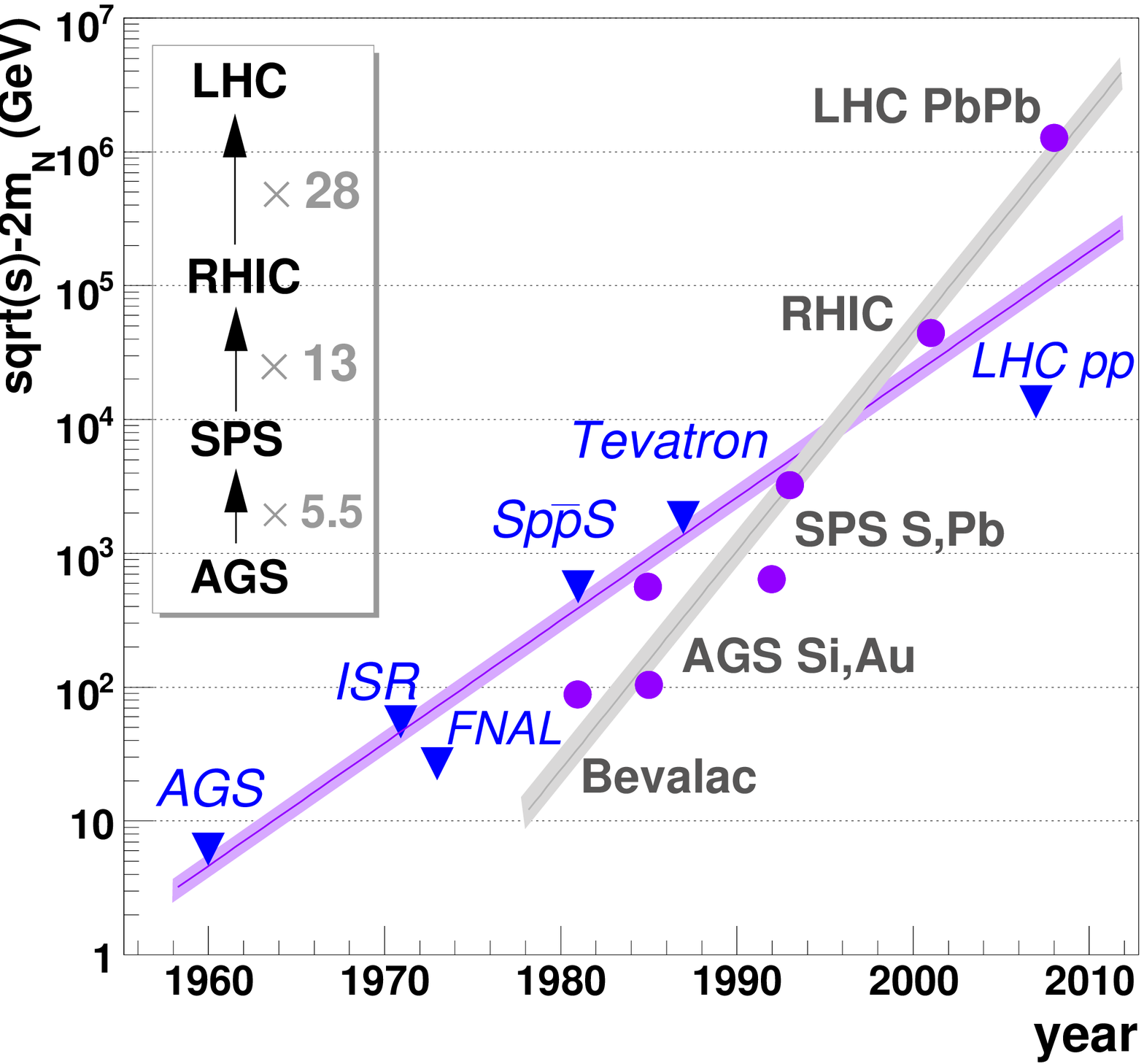}
\hskip 0.25cm
\includegraphics[width=8cm,height=6cm]{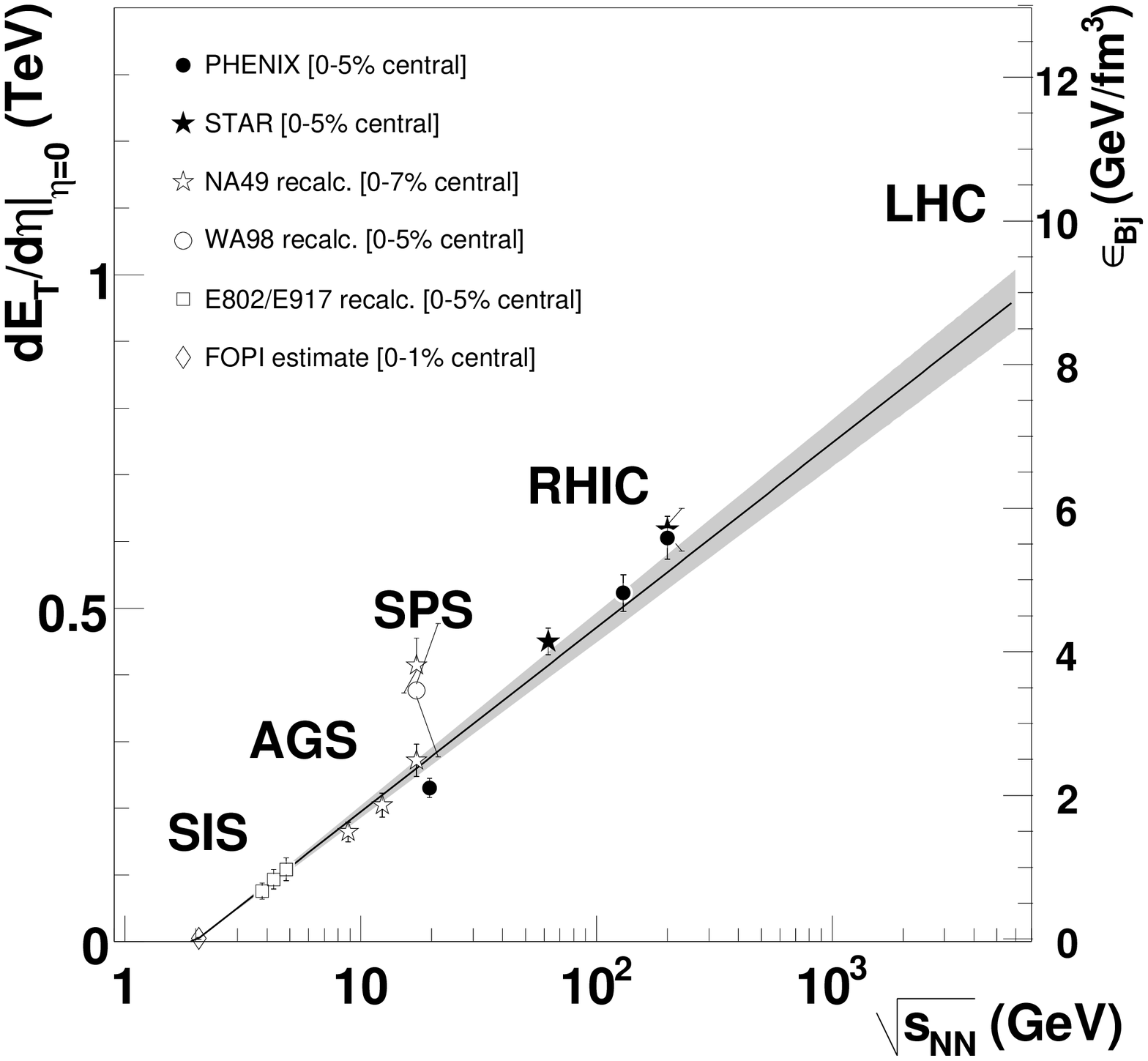}
\vskip -0.20cm
\caption{Left: ``Livingston plot'' for (anti)proton and ion accelerators in the period 1960-2008 
(adapted from~\protect\cite{schukraft06}). Right: Measured transverse energy per unit rapidity
at $\eta$ = 0, and corresponding Bjorken energy density $\ebj(\tau_0$ = 1 fm/$c)$~\cite{bjorken_scaling}, 
in central heavy-ion collisions at various c.m. energies~\protect\cite{ppg019,star_dETdeta} 
fitted to a logarithmic parametrization.}
\label{fig:livingston_plot}
\end{figure}

\noindent
The only experimental means available so far to investigate the (thermo)dynamics of a multi-parton system
involves the use of large atomic nuclei collided at ultrarelativistic energies.  Figure~\ref{fig:livingston_plot} left, 
shows the total center-of-mass energy available for particle production (i.e. subtracting the rest mass of the 
colliding hadrons) at different accelerators as a function of the first operation year (``Livingston plot'')~\cite{schukraft06}. 
The exponential increase in performance translates into an energy doubling every 2 (3) years for the ion
($\bar{p},p$) beams. %, commensurate with the Moore's law for computing power. 
Head-on collisions of heavy ions ($AA$) can produce extremely hot and dense QCD matter by 
concentrating a substantial amount of energy ($\mathscr{O}$(1 TeV) at mid-rapidities at the LHC, 
see Fig.~\ref{fig:livingston_plot} right) in an {\it extended} cylindrical volume 
$V=\pi R_A^2\tau_0\approx$ 150 fm$^3$ for a typical large nucleus with radius $R_A$ = 6.5 fm,
at thermalization times of $\tau_0$ = 1 fm/$c$.

\noindent
The hot and dense systems produced in high-energy $AA$ collisions are not prepared under controlled 
thermodynamical conditions but they follow a dynamical trajectory along the phase diagram shown e.g. in 
Fig.~\ref{fig:QCD_facets}, right. After the collision, the system (with a temperature profile decreasing from 
the center) expands with relativistic longitudinal (transverse) velocities $\mean{\beta}\approx 1.0(0.5)$ and 
cools at rates $T\propto \tau^{-1/n}$ (e.g. $n$ = 3 for a longitudinal-only expansion~\cite{bjorken_scaling}). 
When $T$ reaches $\Tcrit\approx$ 190 MeV, the quark matter undergoes a phase transition into hadrons. 
The produced hadronic gas stops self-interacting collectively at freeze-out times 
$\tau\approx$ 10 -- 20 fm/$c$~\cite{kolb_heinz_rep}. At the initial stages of the reaction 
(1 fm/$c$ after impact), the commonly used ``Bjorken estimate''~\cite{bjorken_scaling} gives energy 
densities attained at mid-rapidity of $\ebj = \dETdeta/(\pi R_A^2\,\tau_0) \approx$ 5, 10 GeV/fm$^3$ 
at RHIC and LHC (Fig.~\ref{fig:livingston_plot} right). Although these values can only be considered as a 
lower limit since they are obtained in a simple 1+1D expansion scenario ignoring any effects from 
longitudinal work, they are already about 5 and 10 times larger, respectively, than the QCD critical energy 
density for deconfinement. High-energy collisions in heavy-ion colliders provide therefore the appropriate 
conditions for the study of highly excited quark-gluon matter.\\
%The first heavy-ion collisions were carried out in the early 80's at the Bevalac accelerator at Berkeley
%up to 2 GeV/nucleon, followed by AGS at BNL and CERN-SPS fixed-target programmes. 
%In year 2000, the Relativistic Heavy-Ion Collider (RHIC) started a extremely fruitful physics plan
%with . When the LHC comes into operation with Pb beams in 2008, the available energy in the 
%center of mass (c.m.) will have increased by four orders of magnitude since the first Bevalac times. 

\noindent
This review is organised as follows. Sections~\ref{sec:probes} introduces the experimental probes of 
subhadronic matter used in high-energy $AA$ collisions. Section~\ref{sec:exp} briefly reviews the
experimental apparatus used in heavy-ion collisions at RHIC and SPS. In Section~\ref{sec:pp_dAu_ref} 
various hard QCD results from proton-proton collisions at RHIC are presented as the ``free space'' baseline 
to which one compares the heavy-ion (``QCD medium'') data. Sections~\ref{sec:saturation}--\ref{sec:chiral} 
each discuss a different physics observable in the context of the latest experimental results available at 
RHIC and SPS. Due to space limitations, the list of topics chosen covers but a fraction of the 
substantial amount of data collected in the last years. %of operation at RHIC. %and SPS. 
%The choice has been mainly driven by 
Those observables that provide direct information on the {\it partonic} phases of the reaction 
have been given preference over ``soft'' or bulk observables from the late hadronic stages. 
Thus, important results relative to the freeze-out phase of the collision -- e.g. chemical and 
kinetic equilibrium from light (and especially strange~\cite{rafelski,antinori04}) hadron abundances~\cite{pbm}, 
``femtoscopy'' measurements from HBT radii~\cite{femtosc}, or possible signatures of (prehadronic)
divergent susceptibilities from final net charge and $\mean{p_T}$ fluctuations~\cite{fluct} -- are not treated 
here. The selected experimental results are mostly from the comprehensive reviews of the four 
RHIC experiments (PHENIX~\cite{phenix_wp}, STAR~\cite{star_wp}, PHOBOS~\cite{phobos_wp}, 
and BRAHMS~\cite{brahms_wp})  from $AuAu$, $dAu$ and $pp$ collisions up to a maximum 
center-of-mass energy of $\sqrtsnn$ = 200 GeV, as well as recent results from NA60 $InIn$
reactions at SPS ($\sqrtsnn$ = 17.3 GeV)~\cite{na60_rho}. 

%%%%%%%%%%%%%%%%%%%%%%%%%%%%%%%%%%%%%%%%%%%%%%%%%%%%%%%%%%%%%%%%

\section{Experimental probes of QCD matter}
\label{sec:probes}

Direct information on the thermodynamical and transport properties of the strongly interacting medium 
produced in $AA$ collisions is commonly obtained by comparing the results for a given observable $\Phi_{AA}$ 
to those measured in  proton(deuteron)-nucleus ($p(d)A$, ``cold QCD matter'') and in proton-proton ($pp$, 
``QCD vacuum'') collisions as a function of c.m. energy, transverse momentum, rapidity, reaction centrality 
(impact parameter $b$), and particle type (mass). Schematically:
\begin{eqnarray}
\!\!\!\!\!\!\!\!\!\!\!\!\!\!\!\!\!\!R_{AA}(\sqrtsnn,p_T,y,m;b) &=&\frac{\mbox{\small{``hot/dense QCD medium''}}}{\mbox{\small{``QCD vacuum''}}}
\, \propto \,\frac{\Phi_{AA}(\sqrtsnn,p_T,y,m;b)}{\Phi_{pp}(\sqrts,p_T,y,m)}\\%\nonumber\\
\!\!\!\!\!\!\!\!\!\!\!\!\!\!\!\!\!\!R_{p(d)A}(\sqrtsnn,p_T,y,m;b) &=&\frac{\mbox{\small{``cold QCD medium''}}}{\mbox{\small{``QCD vacuum'''}}}
\,\propto \,\frac{\Phi_{p(d)A}(\sqrtsnn,p_T,y,m;b)}{\Phi_{pp}(\sqrtsnn,p_T,y,m)}%\nonumber
\end{eqnarray}
The observed {\it enhancements} and/or {\it suppressions} in the $R_{AA,dA}(\sqrtsnn,p_T,y,m;b)$ ratios can 
then be directly linked to the properties of the strongly interacting matter after accounting for a realistic modeling 
of the space-time evolution of the $AA$ expansion process.\\

\noindent
Among the observables whose {\it suppression} is expected to provide information on the produced system,
we will discuss:
\begin{itemize}
\item The {\bf total particle multiplicity} which, related via local parton-hadron duality~\cite{lphd} to the 
number of initially produced partons, will be suppressed if the initial parton flux is reduced due to low-$x$ 
saturation effects in the colliding nuclei~\cite{kharzeev} (Section~\ref{sec:saturation}).
\item {\bf High-$p_T$ leading hadrons} are expected to be produced in reduced yields due to medium-induced
energy loss via gluonsstrahlung of the parent partons in a system with a large number density of colour charges~\cite{bdmps,jet_quench_review} 
(Sections~\ref{sec:jet_quench} and \ref{sec:dijets}).
\item Dissociation of the {\bf heavy quarkonia} bound states has long since proposed~\cite{matsui_satz} 
as a sensitive signature of Debye screening effects above $\Tcrit$ (Section~\ref{sec:jpsi}).
\item The {\bf mass of light vector mesons} is expected to drop in some scenarios of chiral symmetry 
restoration~\cite{chiral_symm,brown_rho} which directly link the in-medium meson mass to a 
temperature- (and baryon density-) dependent chiral condensate: 
$m_V^\star \propto m_V\cdot\mean{\bar{q}q}(T)$ (Section~\ref{sec:chiral}).
\end{itemize}
Likewise, the following observables discussed hereafter are expected to be {\it enhanced} in a strongly-interacting
multi-parton system compared to the results in $pp$ collisions:
\begin{itemize}
\item %{\bf Hydrodynamical radial and elliptical flows}, indicative of the early development of partonic pressure gradients, 
The {\bf soft hadron spectra} ($p_T<$ 2 GeV/$c$), both inclusive and relative to the reaction plane, 
are expected to be boosted due to the development of collective radial and elliptical (hydrodynamical) flows in 
the early partonic phases of the reaction~\cite{kolb_heinz_rep,huovinen_ruuskanen06} (Section~\ref{sec:sQGP}).
\item Semihard ($p_T\approx$ 2--5 GeV/$c$) {\bf yields and flows of baryons} are predicted to be enhanced in 
the context of parton recombination (or coalescence) models~\cite{reco} which take into account modifications of the
hadronization process in a dense deconfined medium (Section~\ref{sec:recomb}).
\item The measurement in $AA$ collisions of a {\bf thermal photon} excess over the expected prompt $\gamma$ 
yield~\cite{stankus05}  %from a thermalized radiating source, 
would provide direct access to the thermodynamical properties (temperature, EoS) of the produced system (Section~\ref{sec:photons}).
\end{itemize}

\begin{figure}[htbp]
\centering
\hskip -0.5cm
\includegraphics[width=8cm,height=6.5cm]{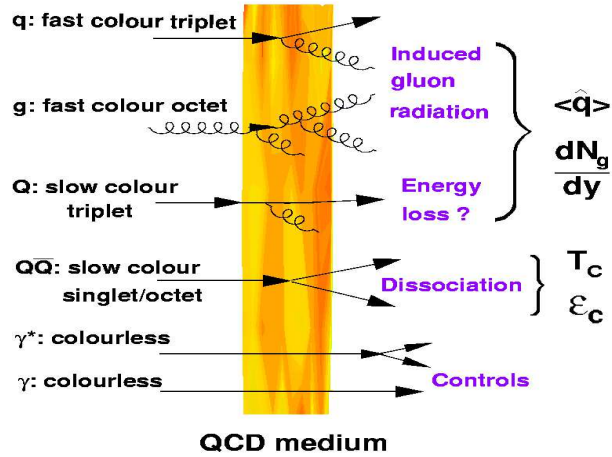}
\vskip -0.2cm
\caption{Examples of hard probes whose modifications in high-energy $AA$ collisions provide 
direct information on properties of QCD matter such as the $\qhat$ transport coefficient, the
initial gluon rapidity density $dN^g/dy$, and the critical temperature and energy density.}
\label{fig:qgp_probes}
\end{figure} 

\noindent
Among all available experimental observables,  particles with large transverse momentum $p_T$ 
and/or high mass (``hard probes'')~\cite{yr_hardprobes_lhc,jacobs_wang} are of  crucial importance for 
several reasons (Fig.~\ref{fig:qgp_probes}): 
(i) they originate from partonic scatterings with large momentum transfer $Q^2$ 
and thus are directly coupled to the fundamental QCD degrees of freedom; 
(ii) their production time-scale is very short, $\tau\approx 1/p_T\lesssim $ 0.1 fm/$c$, 
allowing them to propagate through (and be potentially affected by) the medium, 
(iii) their cross sections can be theoretically predicted using the perturbative QCD 
(pQCD) framework. Hard processes thus constitute experimentally- and theoretically-controlled 
self-generated ``tomographic'' probes of the hottest and densest phases of the reaction.
%inclusive high $p_T$ hadrons, jets, direct photons, Drell-Yan, and heavy flavours, have long been 
%considered both experimentally and theosretically sensitive and well calibrated 
%probes of the small-disHard particle production at high $p_{T}$ ($p_{T}\gtrsim$ 2 GeV/$c$) results from incoherent 
The pQCD {\it factorization theorem}~\cite{factor} allows one to determine the production 
cross section of a given hard probe as the convolution of long-distance parton distribution  
(PDFs, $f_{a/A}$) and fragmentation (FFs, $D_{c\rightarrow h}$) functions and the 
(perturbatively computable up to a given order in $\alpha_s$) parton-parton scattering cross section:
\begin{equation}
d\sigma^{hard}_{AB\rightarrow h}= f_{a/A}(x,Q^2)\otimes f_{b/B}(x,Q^2)\otimes 
d\sigma_{ab\rightarrow c}^{hard} \otimes D_{c\rightarrow h}(z,Q^2)+ \mathscr{O}(1/Q^2)
\label{eq:factorization}
\end{equation}
The validity of Eq.~(\ref{eq:factorization}) holds on the possibility to separate long- and short-distance 
effects with independent QCD time (length) scales
%The possibility to separate long- and short-distance effects in Eq.~(\ref{eq:factorization}) stands on
%the mutual independence of QCD dynamics at different time (length) scales 
as well as on the assumption of {\it incoherent} parton-parton scatterings. In $AA$ collisions, the 
incoherence condition for hard processes implies: $f_{a/A}\approx A\cdot f_{a/N}$,  i.e. that, in the {\it absence} 
of medium effects, the parton flux in a nucleus $A$ should be the same as that of a superposition of $A$ 
independent nucleons. Thus,
\begin{equation}
d\sigma^{hard}_{AB\rightarrow h} \approx A\cdot B\;\cdot f_{a/p}(x,Q^2)\otimes \;f_{b/p}(x,Q^2)\otimes 
d\sigma_{ab\rightarrow c}^{hard}\otimes D_{c\rightarrow h}(z,Q^2),
\label{eq:factorization2}
\end{equation}
%Accordingly, %interaction amplitudes for hard processes (with small cross-sections) add incoherently 
and minimum-bias hard cross sections in $AB$ collisions are expected to scale simply as
%with the number of individual scattering centers:
$d\sigma^{hard}_{AA}|_{{\mbox{\tiny $MB$}}} = A\cdot B\cdot d\sigma_{pp}^{hard}$.
In the most general case, for a given $AB$ reaction with arbitrary impact parameter $b$
the yield can be obtained by multiplying the cross sections measured in $pp$ collisions 
with the ratio of the incident parton flux of the two nuclei:
$dN^{hard}_{AB}(b) = \langle T_{AB}(b)\rangle \cdot d\sigma_{pp}^{hard}$, 
where $T_{AB}(b)$ (normalised to $A\cdot B$) is the nuclear overlap function at $b$ determined
within a purely geometric Glauber eikonal model using the measured Woods-Saxon distribution for 
the colliding nuclei~\cite{dde_glauber}. The standard method to quantify the effects of the medium on the 
yield of a hard probe produced in a $AA$ reaction is thus given by the {\it nuclear modification factor}:
\begin{equation} 
R_{AA}(p_{T},y;b)\,=\,\frac{d^2N_{AA}/dy dp_{T}}{\langle T_{AA}(b)\rangle\,\times\, d^2 \sigma_{pp}/dy dp_{T}},
\label{eq:R_AA}
\end{equation}
which measures the deviation of $AA$ at %impact parameter 
$b$ from an incoherent superposition of $NN$ collisions.
This normalization is usually known as ``binary collision scaling''.

%%%%%%%%%%%%%%%%%%%%%%%%%%%%%%%%%%%%%%%%%%%%%%%%%%%%%%%%%%%%%%%%

\section{Experiments in high-energy heavy-ions physics}
\label{sec:exp}

\begin{figure}[htb]
\centering
\includegraphics[width=7.5cm,height=5.8cm]{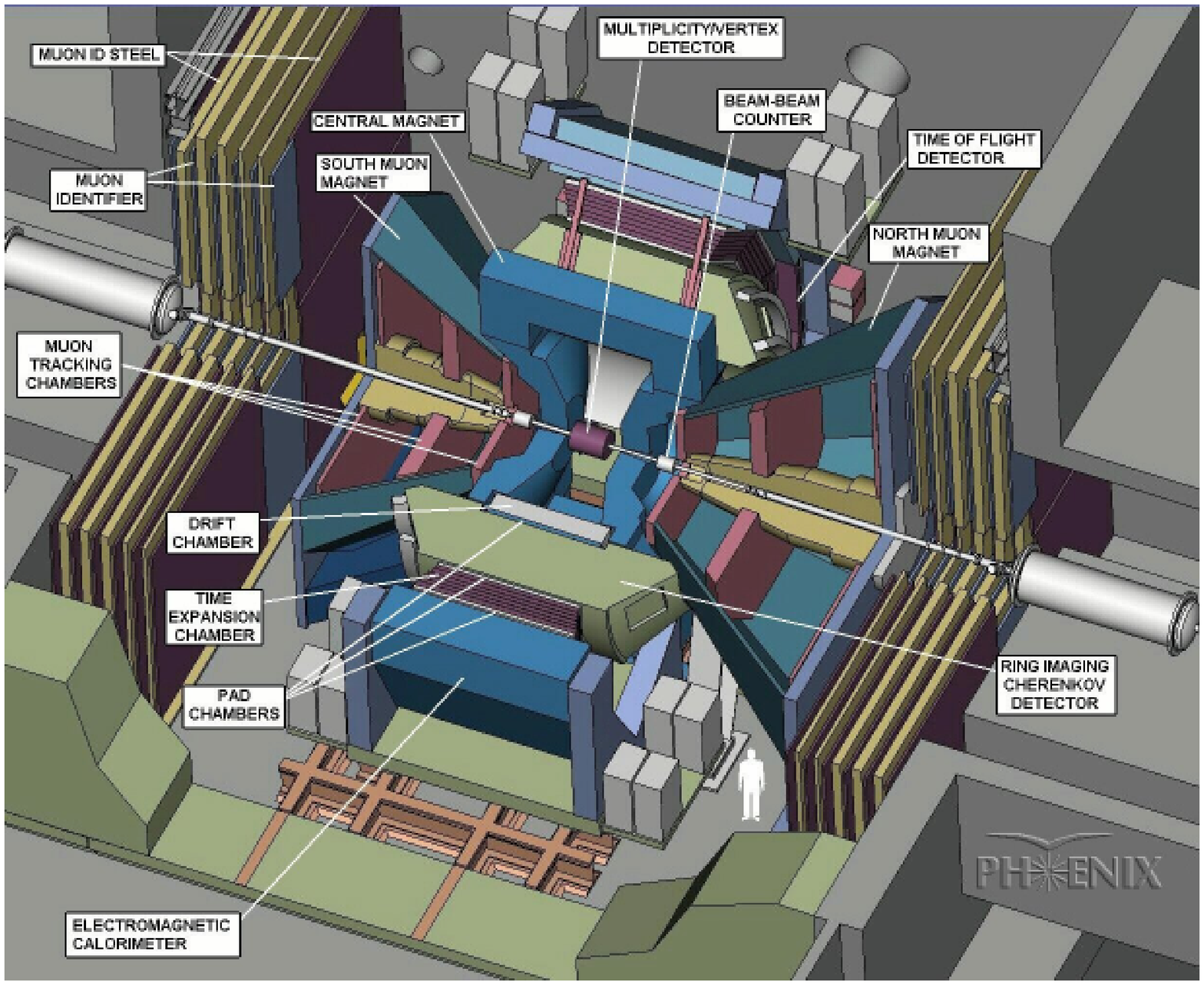}
\includegraphics[width=7.9cm,height=5.8cm]{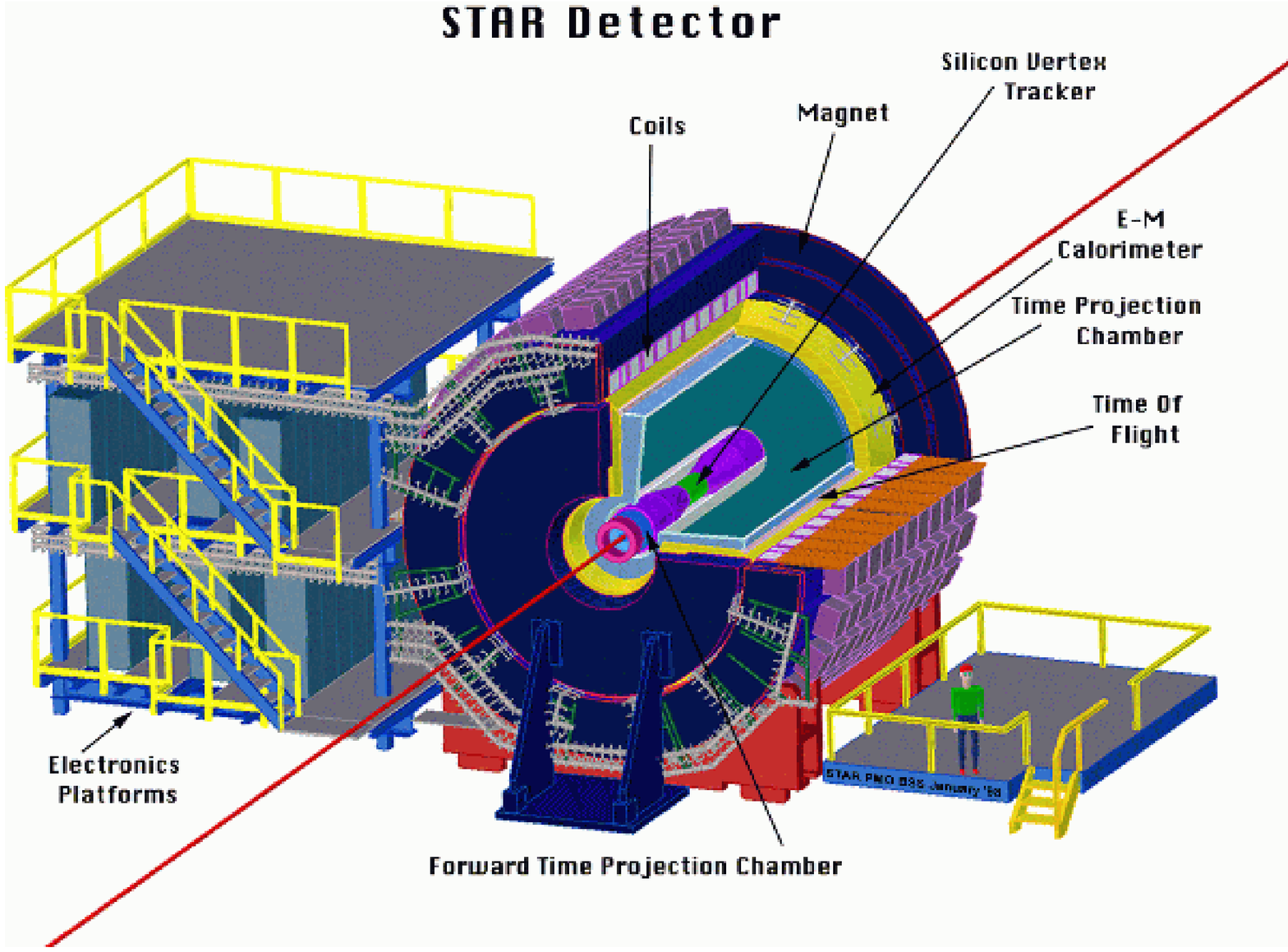}
%\vskip -0.7cm
\caption{The two large experiments at RHIC: PHENIX~\cite{phenix_nim} (left) and STAR~\cite{star_nim} (right).}
\label{fig:star_phenix}
\end{figure}

The Relativistic Heavy Ion Collider (RHIC)~\cite{rhic_nim} at Brookhaven National Laboratory 
is a 3.8-km circumference accelerator composed of two identical, quasi-circular rings of 
superconducting magnets ($\sim$400 dipoles and $\sim$500 quadrupoles) with six crossing-points. 
The machine, which started operation in 1999, can accelerate nuclei (protons) up to a maximum of 
100 (250) GeV/$c$ per nucleon. The center-of-mass energies in $AA$ and $p(d)A$ collisions, $\sqrtsnn$ = 200 GeV 
($\sqrts$ = 500 GeV for $pp$), are more than an order of magnitude larger than those at the 
CERN SPS ($\sqrtsnn$ = 17.3 GeV). The currently attained average $AA$ luminosities\footnote{Note that the %corresponding
``equivalent-$pp$ luminosity'' for hard processes, obtained scaling by the number of  nucleon-nucleon collisions, 
is a much larger value: $\mean{{\cal L}}_{pp-equiv} = A^2\cdot \mean{{\cal L}}_{AA}$ = 
15 $\mu$b$^{-1}$s$^{-1}$.}, $\mean{{\cal L}}\approx 4\times 10^{26}$ cm$^{-2}$s$^{-1}$ = 0.4 mb$^{-1}$s$^{-1}$
are twice the design luminosity thanks mainly to an improved vacuum system~\cite{rhic_epac}.
There are four dedicated experiments at RHIC: two large multi-detector systems (PHENIX and STAR, 
Fig.~\ref{fig:star_phenix}) and two smaller specialised spectrometres (BRAHMS and PHOBOS).\\

\noindent
BRAHMS~\cite{brahms_nim} has two movable %small-acceptance 
magnetic spectrometer arms with hadron identification ($\pi, K, p$) capabilities up to very large rapidities 
($y_{max}\approx$ 4  for charged pions). The two spectrometers consist of a total of 5 tracking chambers 
and 5 dipole magnets (maximum field of 1.7 T), plus 2 Time-Of-Flight systems and a Ring Imaging 
\v{C}erenkov Detector (RICH) for particle identification. The angular coverage of the forward spectrometer 
(FS) goes from 2.3$^\circ$ $<\theta<$ 30$^\circ$ (solid angle of 0.8 msr) up to momenta of 35 GeV/$c$. 
The midrapidity spectrometer (MSR) covers 30$^\circ$ $<\theta<$ 95$^\circ$ (acceptance 6.5 msr).
%Triggering and global even characterization of heavy ion events are carried out in the mid-rapidity multiplicity array, 
%forward scintillator detectors, and the ZDCs.
The PHOBOS experiment~\cite{phobos_nim} is based %almost entirely 
on silicon pad detectors and covers nearly the full solid angle (11 units of pseudorapidity) 
for charged particles, featuring excellent global event characterization at RHIC. PHOBOS consists of 4 subsystems: 
a multiplicity array (``octagon'', $|\eta|<$3.2, and ``rings'', $|\eta|\lesssim$5.4), a finely segmented 
vertex detector, a two-arm magnetic spectrometer (with dipole magnet of strength 1.5 Tm) at midrapidity 
including a time-of-flight wall, and several trigger detectors. 
%The spectrometer arms use a warm dipole magnet of strength 1.5 Tm. 
Particle identification (PID) is based on time-of-flight and energy loss in the silicon.\\
%Special emphasis is put on measurements at very low transverse 
%momentum, requiring a thin Be beam pipe and minimal material in front of the first tracking planes.\\

\noindent
PHENIX~\cite{phenix_nim} is a high interaction-rate experiment specifically designed to measure 
hard QCD probes such as high-$p_T$ hadrons, direct $\gamma$, lepton pairs, and heavy flavour.
PHENIX (Fig.~\ref{fig:star_phenix}, left) achieves good mass and PID resolution, and small granularity 
by combining 13 detector subsystems ($\sim$350,000 channels) divided into: (i) 2 central arm 
spectrometers for electron, photon and hadron measurement at mid-rapidity ($|\eta|<0.35$, 
$\Delta\phi=\pi$); (ii) 2 forward-backward spectrometers for muon detection 
($|\eta|$ = 1.15 - 2.25, $\Delta\phi = 2\pi$); and (iii) 4 global (inner) detectors for trigger 
and centrality selection. Two types of electromagnetic calorimeters, PbSc and PbGl,
measure $\gamma$ and $e^\pm$. Additional electron identification is possible thanks to the 
RICH detector. Charged hadrons are measured in the axial central magnetic field (strength 1.15 Tm) 
by a drift chamber (DC) and 3 layers of MWPC's with pad readout (PC). Hadron identification 
($\pi^\pm$, $K^\pm$, and $p,\bar{p}$) is achieved by matching the reconstructed tracks 
to hits in a time-of-flight wall (TOF).  Triggering is based on two Beam-Beam Counters 
(BBC, $|\eta|$ = 3.0 - 3.9) and the Zero Degree Calorimeters (ZDC, with $|\theta|<$ 2 mrad). 
PHENIX features a state-of-the-art DAQ  system capable of recording $\sim$300 MB/s to disk with 
event sizes of  $\sim$100 KB (i.e. coping with event rates of $\sim$3 kHz).\\

\noindent
The STAR experiment~\cite{star_nim} is based around a large-acceptance Time Projection 
Chamber (TPC) inside a solenoidal magnet with radius 260 cm and maximum field strength 
0.5 T (Fig.~\ref{fig:star_phenix}, right). The TPC  with radius 200 cm and full azimuthal 
acceptance over $|\eta|<$1.4 provides exceptional charged particle tracking and PID via ionization 
energy loss $dE/dx$ in the TPC gas and reconstruction of secondary vertices for weakly decaying 
particles. Additional tracking is provided by inner silicon drift detectors at midrapidity and forward 
radial-drift TPCs at 2.5$<|\eta|<$4. Photons and electrons are measured in the Barrel and Endcap 
Electromagnetic Calorimeters (EMC) covering -1.0$<\eta<$2.0 and full $\phi$.
The large STAR coverage allows for multi-particle correlation studies, jet reconstruction in
$pp$, and also measurement of strange and charm hadrons. Triggering is done with 
the ZDCs, forward scintillators (Beam-Beam counters), a barrel of scintillator slats 
surrounding the TPC, and the EMC.\\

\noindent
NA60~\cite{na60exp} is a high interaction-rate {\it fixed-target} experiment 
focused on the study of dimuon, vector meson and open charm production 
in $pA$ and $AA$ collisions at the CERN SPS. The 17~m long muon
spectrometer previously used by NA38 and NA50, is composed of 8 multi-wire
proportional chambers, 4 scintillator trigger hodoscopes, and a toroidal
magnet.  This spectrometer is separated by a $\sim$\,5~m long hadron
absorber (mostly carbon) from the vertex region, where a silicon tracker
made of pixel detectors~\cite{na60_pixels} placed in a 2.5~T dipole
field, measures the produced charged particles. The matching between the 
muons and the vertex tracks leads to an excellent dimuon mass resolution.

%%%%%%%%%%%%%%%%%%%%%%%%%%%%%%%%%%%%%%%%%%%%%%%%%%%%%%%%%%%%%%%%

\section{Benchmark channels in the vacuum and in cold QCD matter}
\label{sec:pp_dAu_ref}

\begin{figure}[htb]
\centering
\includegraphics[width=8.6cm,height=8.cm]{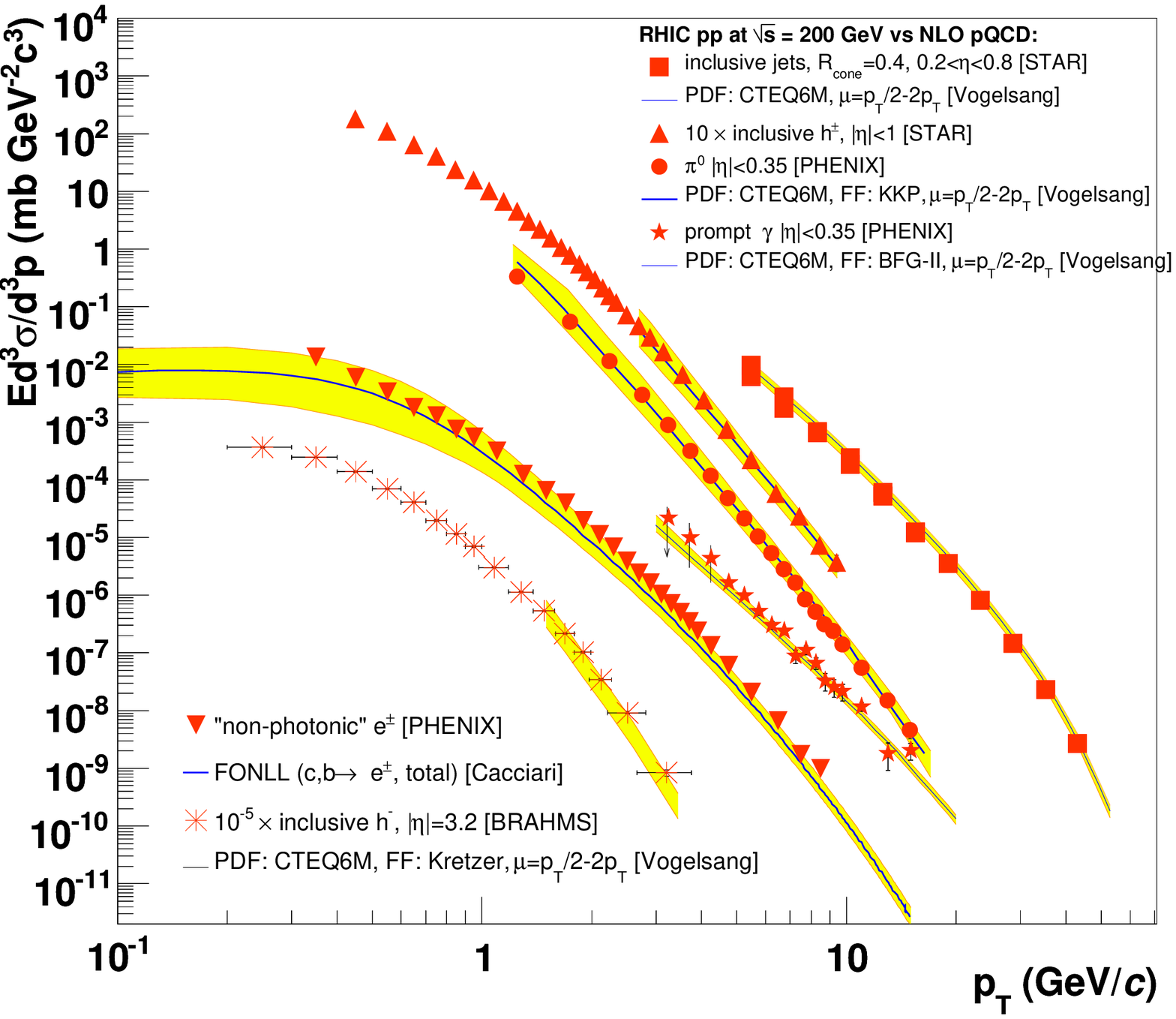}
\includegraphics[width=6.9cm,height=8.cm]{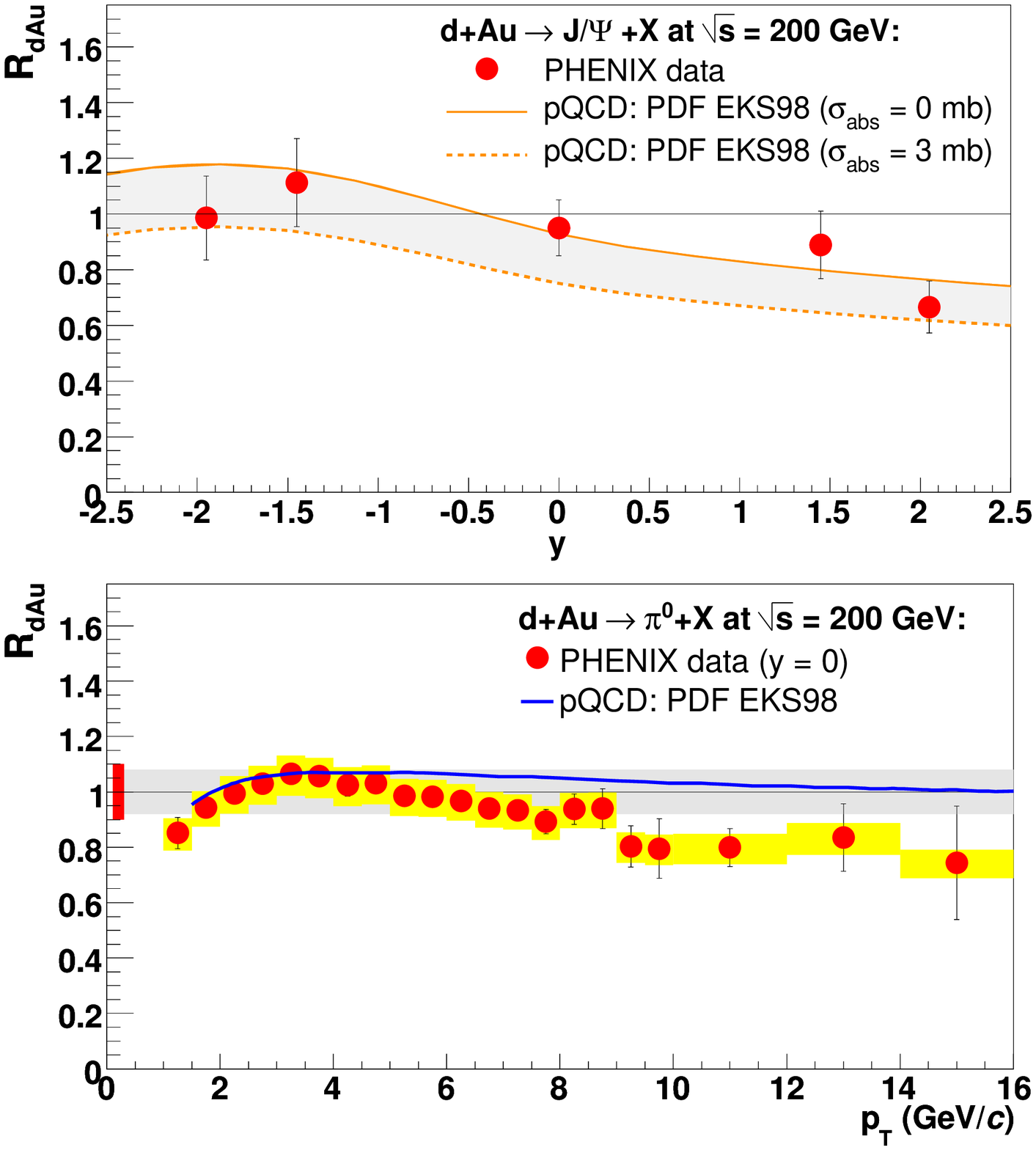}
%\vskip -0.7cm
\caption{Left: Compilation of hard cross sections in $pp$ collisions at $\sqrts$ = 200 GeV (10\%-30\% 
syst. uncertainties not shown  for clarity purposes) measured
by STAR~\cite{star_pp_jets_200,star_hipt_200}, 
PHENIX~\cite{phnx_pp_pi0_200,phnx_pp_gamma_200,phnx_pp_nonphot_elec_200}, 
and BRAHMS~\cite{brahms_pp_dAu} compared to 
NLO~\cite{vogelsang_pi0,vogelsang_gamma} and NLL~\cite{cacciari05} pQCD predictions (yellow
bands). Right: Nuclear modification factors for $\jpsi$ versus rapidity~\cite{phnx_dAu_jpsi} (top), 
and for high-$p_T$ $\pi^{\circ}$ at mid-rapidity~\cite{phnx_dAu_pi0} (bottom) in central $dAu$ 
collisions at $\sqrtsnn$ = 200 GeV compared to pQCD calculations~\cite{vogt05,guzey04} with 
EKS98~\cite{eks98} nuclear shadowing.}
\label{fig:pp_rhic_vs_nlo}
\end{figure}

Proton-proton collisions are the baseline free-space reference to which one compares the $AA$ 
results in order to identify initial- or final-state effects which modify the expectation of equation 
(\ref{eq:factorization2}) and can thus provide direct information on the underlying QCD medium. 
%Modifications of different hard cross-sections in $AA$ collisions are directly related to the 
%medium properties (Fig.~\ref{fig:qgp_probes}).
Figure~\ref{fig:pp_rhic_vs_nlo} collects six different $p_T$-differential inclusive cross sections
measured  at RHIC in $pp$ collisions at $\sqrts$ = 200 GeV: jets~\cite{star_pp_jets_200},
charged hadrons~\cite{star_hipt_200}, neutral pions~\cite{phnx_pp_pi0_200},
direct photons~\cite{phnx_pp_gamma_200}, $D, B$ mesons (indirectly measured via $e^\pm$
from their semileptonic decay)~\cite{phnx_pp_nonphot_elec_200} all at central rapidities ($y=0$), and 
negative hadrons at forward rapidities ($\eta$ = 3.2)~\cite{brahms_pp_dAu}. The measurements
cover 9 orders of magnitude in cross section (from 10 mb down to 10 pb), and broad ranges
in transverse momentum (from zero momentum for $D,B$ mesons up to $\sim$45 GeV/$c$, a half
of the kinematical limit, for jets) and rapidity (from $\eta$ = 0 up to  $\eta$ = 3.2 for $h^-$
and even, not shown here, $y$ = 3.8 for $\pi^{\circ}$'s~\cite{star_fwd_pp_pi0}).
Whenever there is a concurrent measurement of the same observable by two or more RHIC experiments, 
the  data are consistent with each other (the only exception being the heavy-flavour electron 
cross sections, which are a factor 2-3 larger in STAR~\cite{star_nonphoton_elect_AuAupp200}
than in PHENIX~\cite{phnx_pp_nonphot_elec_200}).

\noindent
Standard next-to-leading-order (NLO)~\cite{vogelsang_pi0,vogelsang_gamma} or resummed 
next-to-leading log (NLL)~\cite{cacciari05} perturbative QCD calculations with modern proton 
PDFs~\cite{cteq}, hadron fragmentation functions~\cite{kkp,kretzer}, and with varying 
factorization-renormalization scales ($\mu=p_T/2-2p_T$) show an overall good agreement with 
the available $pp$ data at $\sqrt{s}$ = 200 GeV (yellow bands in Fig. \ref{fig:pp_rhic_vs_nlo}, left). 
This is true even in the semi-hard range $p_T\approx 1- 4$ GeV/$c$, where a perturbative description 
would be expected to give a poorer description of the spectra. These results indicate that the hard QCD 
cross sections at RHIC energies are well under control both experimentally and theoretically in their full
kinematic domain. This is at variance with measurements at lower (fixed-target) energies where several 
data-theory discrepancies still remain~\cite{bourre,aurenche06}.\\

\noindent
Not only the proton-proton hard cross sections are well under control theoretically at RHIC but the 
hard yields in deuteron-nucleus collisions do not show any significant deviation from the perturbative 
expectations. Figure~\ref{fig:pp_rhic_vs_nlo} right, shows  the nuclear modification factors measured in 
$dAu$ collisions at $\sqrtsnn$ = 200 GeV for $\jpsi$, $R_{dAu}(y)$~\cite{phnx_dAu_jpsi}, and for 
leading $\pi^{\circ}$, $R_{dAu}(p_T)$ at $y=0$~\cite{phnx_dAu_pi0}. At {\it mid-rapidity}, the maximum 
deviation from the $R_{dAu}$ = 1 expectation for hard processes without initial-state effects is of the 
order of  $\sim$20\%. Both $R_{dAu}(y,p_T)$ ratios are well accounted for by standard pQCD 
calculations~\cite{vogt05,guzey04} that include DGLAP-based parametrizations of nuclear shadowing~\cite{eks98}
and/or a mild amount of initial-state $p_T$ broadening~\cite{yr_lhc_pdfs} to account for a modest 
``Cronin enhancement''~\cite{cronin}. %and/or a small amount of nuclear absorption. 
These data clearly confirm that at mid rapidities, the parton flux of the incident gold nucleus can be basically 
obtained by geometric superposition of the nucleon PDFs, and that the nuclear $(x,Q^2)$ modifications 
of the PDFs are very modest. %in this kinematic domain. 
Since no final-state dense and hot system is expected to be produced in $dAu$ collisions, such results 
indicate that any deviations from $R_{AA}$ = 1 larger than $\sim$40\% potentially observed for hard probes 
in  $AuAu$ collisions (at central rapidities) can only be due to {\it final}-state effects in the  medium 
produced in the latter reactions.

%%%%%%%%%%%%%%%%%%%%%%%%%%%%%%%%%%%%%%%%%%%%%%%%%%%%%%%%%%%%%%%%

\section{Low-$x$ gluon saturation: $AA$ rapidity densities, and high-$p_T$ $dA$ forward suppression}
\label{sec:saturation}

\begin{figure}[htb]
\centering
\includegraphics[width=6.7cm,height=7cm]{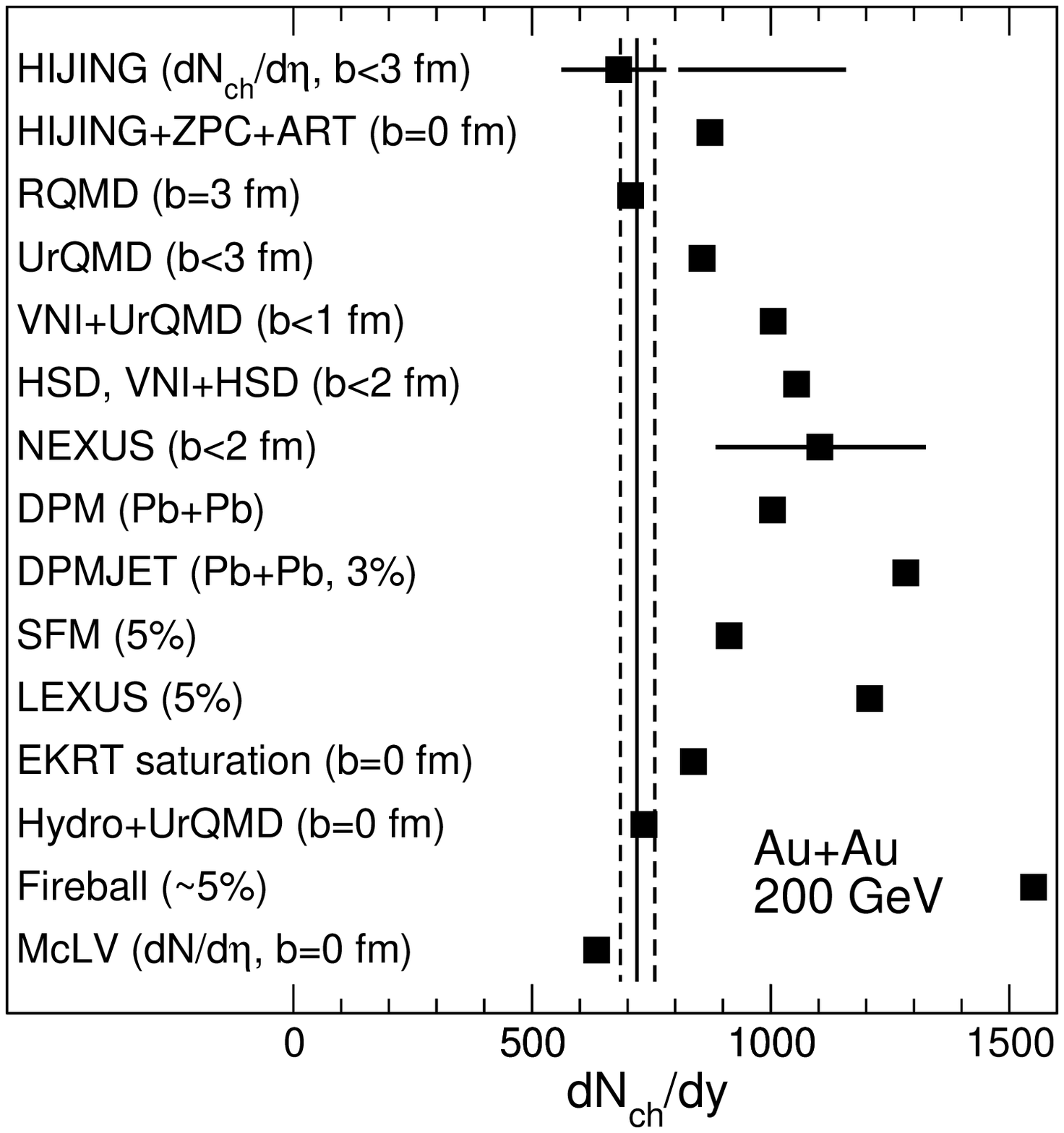}
\includegraphics[width=8.8cm]{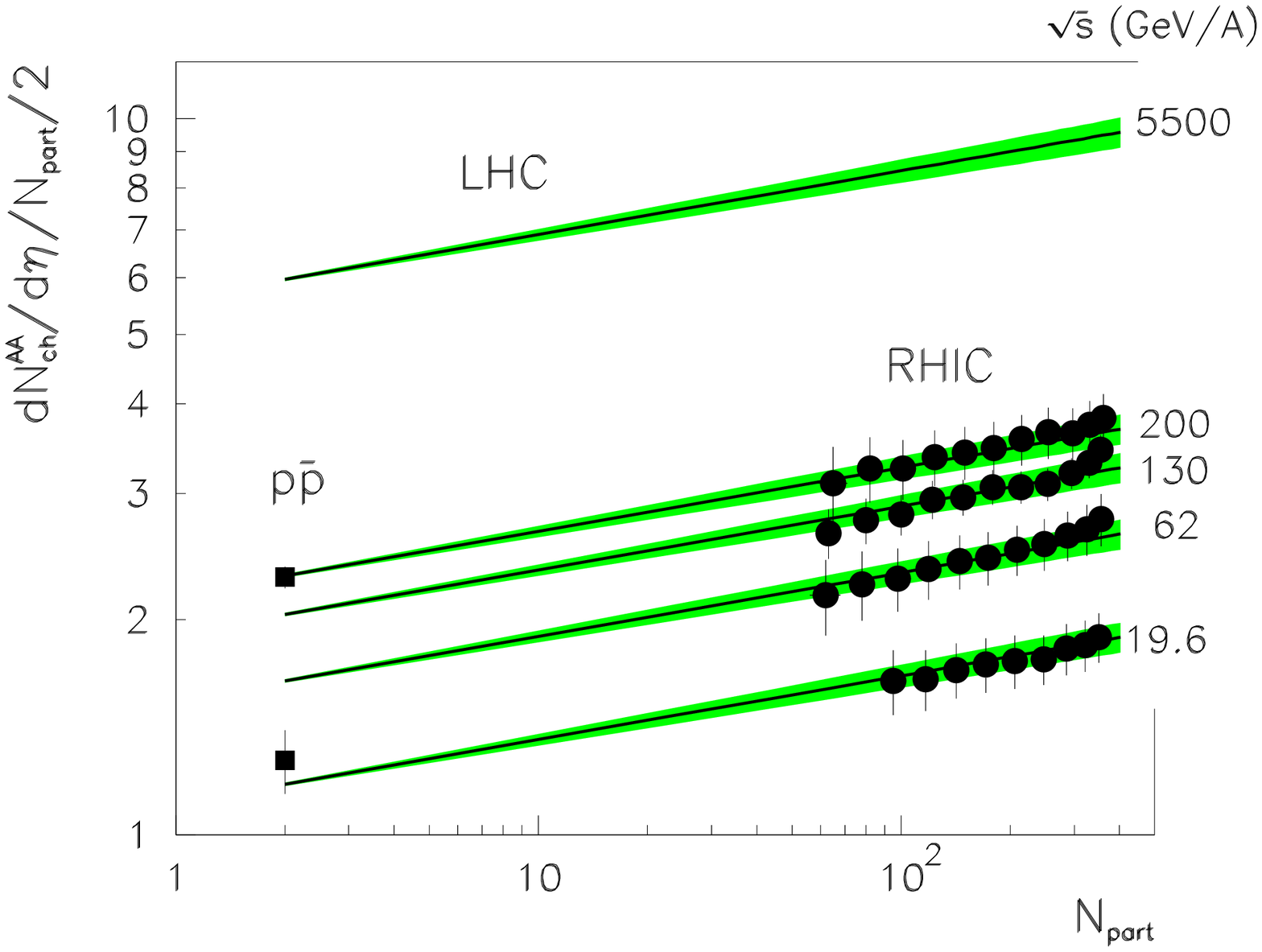}
%\includegraphics[width=7.5cm,height=7cm]{denterria_dNdeta_rhic_predictions.eps}
%\vskip -0.7cm
\caption{Left: Data vs. models for $\dNdeta$ in central $AuAu$ at $\sqrtsnn$ = 200 GeV~\cite{phobos_wp,eskola_qm01} 
(the saturation model prediction is identified as McLV~\cite{mclv}).
Right: Energy- and centrality- (in terms of the number of nucleons participating in the collision, 
$N_{\rm part}$) dependences of  $\dNdeta$  (normalised by $N_{\rm part}$):
%of $(2/N_{\rm part})[dN^{A-A}_{\rm ch}/d\eta](\eta\sim 0)=0.47[\sqrt{s_{_{\ensuremath{\it{NN}}}}}]^{0.288} N_{\rm part}^{0.089}$ 
PHOBOS data~\protect\cite{phobos_wp} versus the saturation prediction~\protect\cite{armesto04} . }
\label{fig:dNdeta}
\end{figure}

The bulk hadron multiplicities measured at mid-rapidity in central $AuAu$ at $\sqrtsnn$ = 200 GeV  
$\dNdeta\approx$ 700, %. Such a particle density is 
are comparatively lower than the $\dNdeta\approx$ 1000 expectations of ``minijet'' dominated 
scenarios~\cite{hijing},  soft Regge models~\cite{dpm} (without accounting for strong shadowing 
effects~\cite{perco}), or extrapolations from an incoherent sum of proton-proton collisions~\cite{eskola_qm01}
(Fig.~\ref{fig:dNdeta}, left). However, Colour Glass Condensate (CGC) approaches~\cite{kharzeev,armesto04} 
which effectively take into account a reduced {\it initial} number of scattering centers in the nuclear PDFs, 
$f_{a/A}(x,Q^2)<A\cdot f_{a/N}(x,Q^2)$, agree well with experimental data. In the saturation 
models non-linear effects become important and start to saturate the parton densities when the area 
occupied by the partons becomes similar to that of the hadron, $\pi R^2$. For a nucleus with $A$ 
nucleons and total gluon distribution $xG(x,Q^2)$ %=A\cdot xg(x,Q^2$ where $g(x,Q^2)$ is the gluon density in a nucleon) 
this condition translates into the following saturation momentum~\cite{GLR,MQ}:
\begin{equation}
Q_s^2(x)\simeq \alpha_s \frac{1}{\pi R^2}\,xG(x,Q^2)\sim A^{1/3}\,x^{-\lambda} \sim A^{1/3}(\sqrts)^{\lambda} \sim A^{1/3}e^{\lambda y},
%Q_s^2(x)\simeq \alpha_s \frac{xG(x,Q^2)}{\pi R^2}\sim \frac{A^{1/3}}{x^\lambda}\sim A^{1/3}(\sqrts)^{\lambda},
%Q_s2 ~ A^{1/3}  s^{lambda/2},with lambda = 0.25, see e.g. eq(17) of hep-ph/0408050
\label{eq:Qs}
\end{equation}
with $\lambda\approx$ 0.2--0.3~\cite{kharzeev}. The mass number dependence implies that, at comparable 
energies, non-linear effects will be $A^{1/3}\approx$ 6 times larger in a heavy nucleus ($A\sim$ 200 for $Au$ or $Pb$) 
than in a proton. Based on the general expression~(\ref{eq:Qs}), CGC-based models can describe the 
centrality and c.m. energy dependences of the bulk $AA$ hadron production (Fig.~\ref{fig:dNdeta}, right).\\

\begin{figure}[htb]
\begin{center}
\includegraphics[width=7.8cm,height=5.cm]{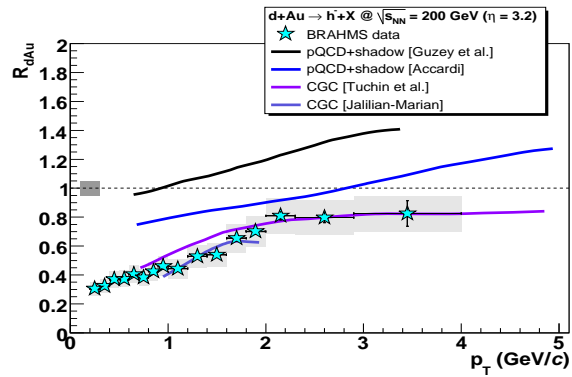}
\caption{Nuclear modification factor $R_{dAu}(p_T)$ for negative hadrons measured at forward
rapidities by BRAHMS in $dAu$  at $\sqrtsnn$ = 200 GeV~\protect\cite{brahms_pp_dAu} compared
to predictions of leading-twist shadowing pQCD~\protect\cite{guzey04,accardi04} (two upper curves)
 and CGC~\protect\cite{tuchin04,jamal04} (lower curves).}
\label{fig:RdA}
\end{center}
\vspace{-.4cm}
\end{figure}

\noindent
The second manifestation of CGC-like effects in the RHIC data is the BRAHMS observation of suppressed 
yields of moderately high-$p_T$ hadrons ($p_T\approx 2 - 4 $ GeV/$c$) in $dAu$ relative to $pp$ 
collisions at forward rapidities ($\eta\approx$ 3.2, Fig.~\ref{fig:RdA})~\cite{brahms_pp_dAu}. 
Hadron production at such small angles is theoretically sensitive to partons in the $Au$ nucleus 
with  $x_2^{min} = (p_T/\sqrtsnn)\,\exp(-\eta)\approx\mathscr{O}$(10$^{-3}$)~\cite{guzey04}.
The observed nuclear modification factor, $R_{dAu}\approx$ 0.8, cannot be reproduced by pQCD
calculations that include the same {\it leading-twist} nuclear shadowing~\cite{guzey04,accardi04,deflorian03}
that describes the $dAu$ data at $y=$ 0 (Fig.~\ref{fig:pp_rhic_vs_nlo}, right) but can be described 
by CGC approaches that parametrise the $Au$ nucleus as a saturated gluon wavefunction~\cite{tuchin04,jamal04,dumitru05}. 

\noindent
It is worth noting, however, that at RHIC energies the saturation momentum is in the transition 
between the soft and hard regimes ($Q_s^2\approx$ 2 GeV$^2$) and that the results consistent 
with the CGC predictions are in a kinematic range with relatively low momentum scales: $\mean{p_T}\sim$ 0.5 GeV
for the bulk hadron multiplicities and  $\mean{p_T}\sim$ 2 GeV for forward inclusive hadron production.
In such a kinematic range non-perturbative effects can blur a simple interpretation based on partonic degrees of 
freedom alone. Following Eq.~(\ref{eq:Qs}), the relevance of low-$x$ QCD effects  will certainly be enhanced 
at the LHC due to the increased: center-of-mass energy, nuclear radius ($A^{1/3}$), and rapidity of the produced 
partons~\cite{yr_lhc_pdfs,dde_lowx06}. At the LHC, the saturation momentum $Q_s^2\approx$ 5 -- 10 GeV$^2$~\cite{kharzeev}
will be more clearly perturbative and the relevant $x$ values in $AA$ and $pA$ collisions  
will be 30--70 times lower than at RHIC: $x_{2}\approx 10^{-3}(10^{-5})$  at central (forward) 
rapidities for processes with a hard scale $p_T\sim$10 GeV. 

%%%%%%%%%%%%%%%%%%%%%%%%%%%%%%%%%%%%%%%%%%%%%%%%%%%%%%%%%%%

\section{sQGP viscosity: Strong hydrodynamical flows and AdS/CFT connection}
\label{sec:sQGP}

\begin{figure}[htb]
\centering
\includegraphics[width=8cm]{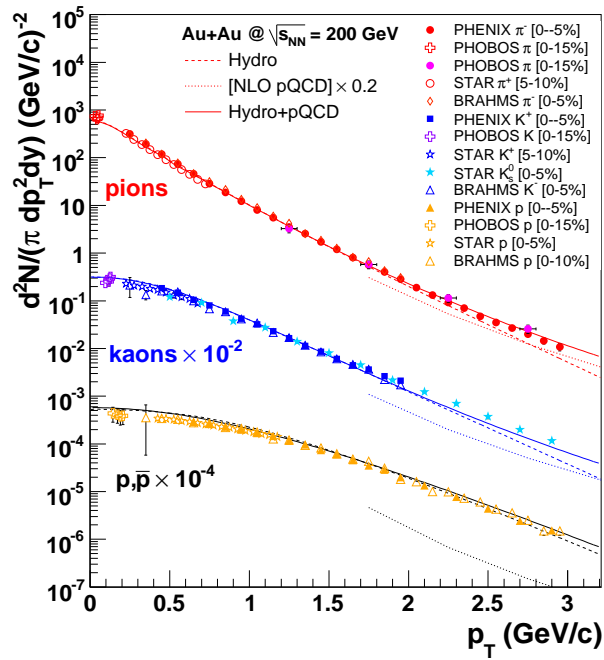}
%\vskip -0.7cm
\caption{Transverse momentum spectra for pions, kaons, and (anti)protons measured at RHIC below 
$p_T\approx$ 3 GeV/$c$ in 0-10\% central $AuAu$ collisions at $\sqrtsnn$ = 200 GeV compared to 
hydrodynamics(+pQCD) calculations~\protect\cite{dde_peressou}.}
\label{fig:hydro_spec}
\end{figure}

The bulk hadron production ($p_T\lesssim$ 2 GeV/$c$) in $AuAu$ reactions at RHIC shows strong 
collective effects known as radial and elliptic flows. First, the measured single hadron $p_T$
spectra have an inverse slope parameter $\Teff$ larger than that measured in $pp$ collisions, increasing
with reaction centrality and with hadron mass as expected if collective expansion blue-shifts
the hadron spectra. Empirically, one finds: $\Teff \approx T + \mean{\beta_T}^2\cdot m$, with 
$T$ and $\mean{\beta_T}$ being the freeze-out temperature and average collective flow velocity
of the ``fireball'', and $m$ the mass of the hadron. Phenomenological fits of the spectra to ``blast wave'' 
models yield transverse flow velocities $\mean{\beta_T}\approx$ 0.6~\cite{phenix_wp}. Full hydrodynamical 
calculations which start with a partonic phase very shortly after impact ($\tau_0<$ 1 fm/$c$) develop 
the amount of collective radial flow needed to accurately reproduce all the measured hadron spectra 
(Fig.~\ref{fig:hydro_spec}).

\begin{figure}[htb]
\centering
\includegraphics[width=7.5cm , bb= -20 -20 366 269]{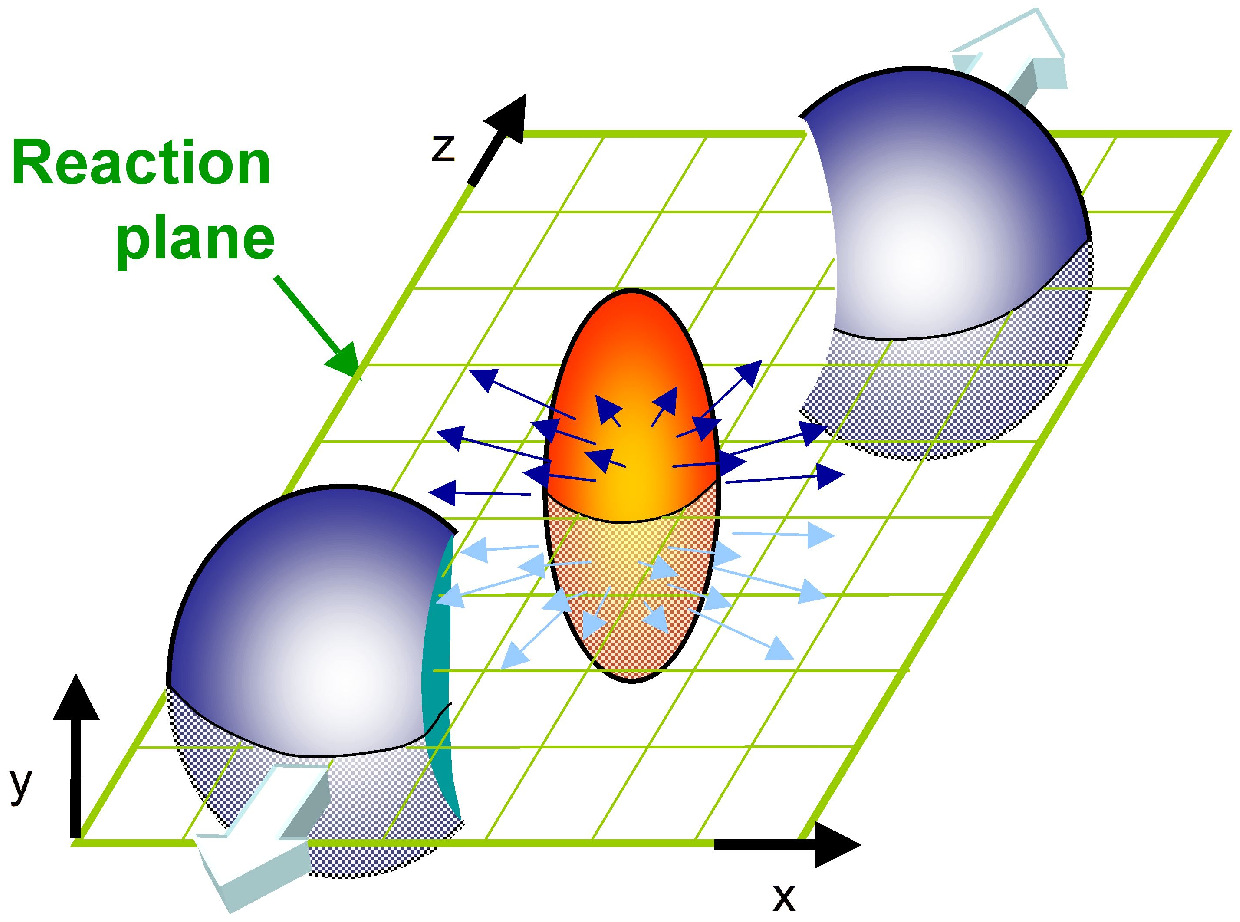}
\includegraphics[width=8cm]{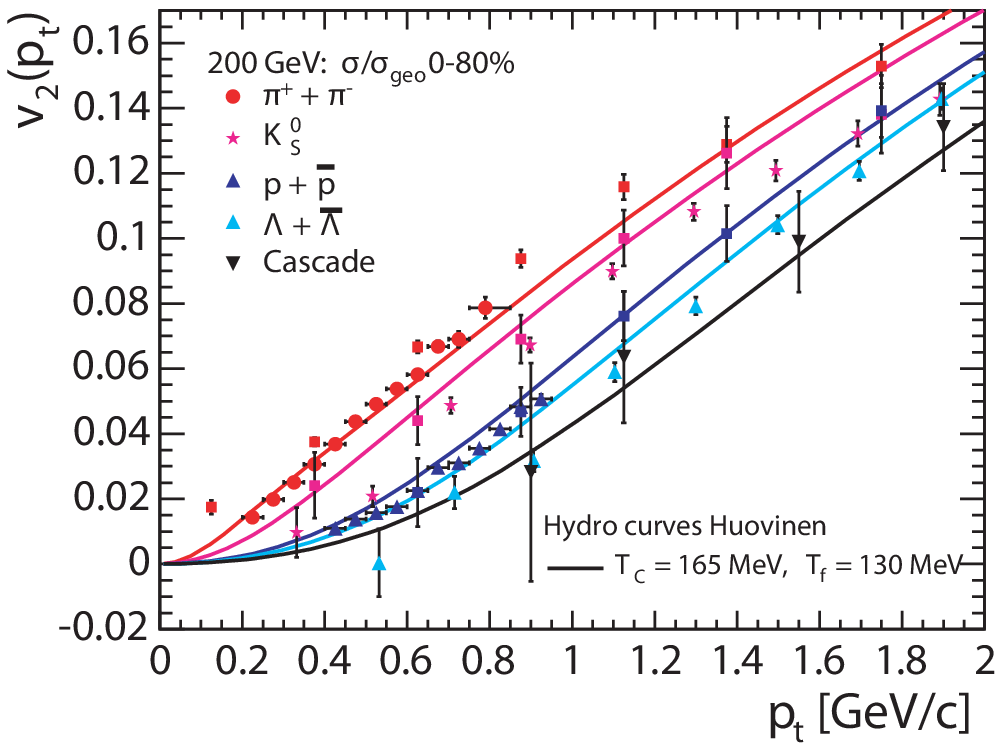}
%\vskip -0.7cm
\caption{Left:  Spatial asymmetry with respect to the reaction plane of the produced ``fireball'' in non-central 
nucleus-nucleus collisions. Right:  Measured elliptic flow parameter $v_2(p_T)$ below $p_T =$ 2 GeV/$c$ 
for a variety of hadrons~\protect\cite{snellings05} compared to hydrodynamic 
predictions~\protect\cite{huovinen_ruuskanen06}.}
\label{fig:v2}
\end{figure}

\noindent
Secondly, the azimuthal distributions $dN/d\Delta\phi$ of hadrons emitted relative to the reaction plane 
($\Delta\phi=\phi-\Phi_{RP}$) show a strong harmonic modulation with a preferential ``in-plane'' emission in 
non-central collisions. Such an azimuthal flow pattern is a truly collective effect (absent in $pp$ collisions) 
consistent with an efficient translation of the initial coordinate-space anisotropy in $AA$ reactions with non-zero 
impact parameter (i.e. with a lens-shaped overlap zone, see Fig.~\ref{fig:v2} left) into a final ``elliptical'' 
asymmetry in momentum-space. Rescattering between the produced particles drives collective motion along 
the pressure gradient, which is larger for directions parallel to the smallest dimension of the lens. The strength 
of this asymmetry is quantified via the second Fourier coefficient $v_2(p_T,y)\equiv\mean{cos(2\Delta\phi)}$ 
of the azimuthal decomposition of single inclusive hadron spectra with respect to the reaction plane~\cite{ollitr92,voloshin94}
\begin{equation}
E\frac{d^3N}{d^3p}=\frac{1}{2\pi}\frac{d^2N}{p_T\,dp_T\,dy}\left(1+2\sum_{n=1}^{\infty}v_n\,cos[n(\phi-\Phi_{RP})]\right).
\end{equation}
The large $v_2\approx$ 0.16 measured in the data (Fig.~\ref{fig:v2}, right) indicates a strong degree of 
collectivity (pressure gradients) building up in the first instants of the collision. Indeed,  elliptic flow 
develops in the initial phase of the reaction and quickly self-quenches beyond $\tau\approx$ 5 fm/$c$ as the 
original spatial eccentricity disappears~\cite{kolb00}. Two additional experimental observations support the existence 
of an efficient hydrodynamical response with very short thermalization times. First, not only light hadrons but 
also $D,B$ mesons (indirectly measured via their semileptonic decays into $e^\pm$) 
show momentum anisotropies with $v_2$ as large as 10\%~\cite{akiba05}. 
The fact that the heavy $c$ and $b$ quarks participate in the common 
flow of the medium is clearly suggestive of a robust collective response during the early {\it partonic} phase. 
In addition, the $v_2$ values measured for different centralities, at different center-of-mass energies 
(200 and 62 GeV) and for different colliding systems ($AuAu$ and $CuCu$) are found to show simple scaling laws 
with the reaction eccentricity~\cite{phobos_qm05,phnx_ppg062} also in agreement with hydrodynamics expectations.\\

\begin{figure}[htb]
\begin{center}
\includegraphics[width=9.5cm,height=6.5cm]{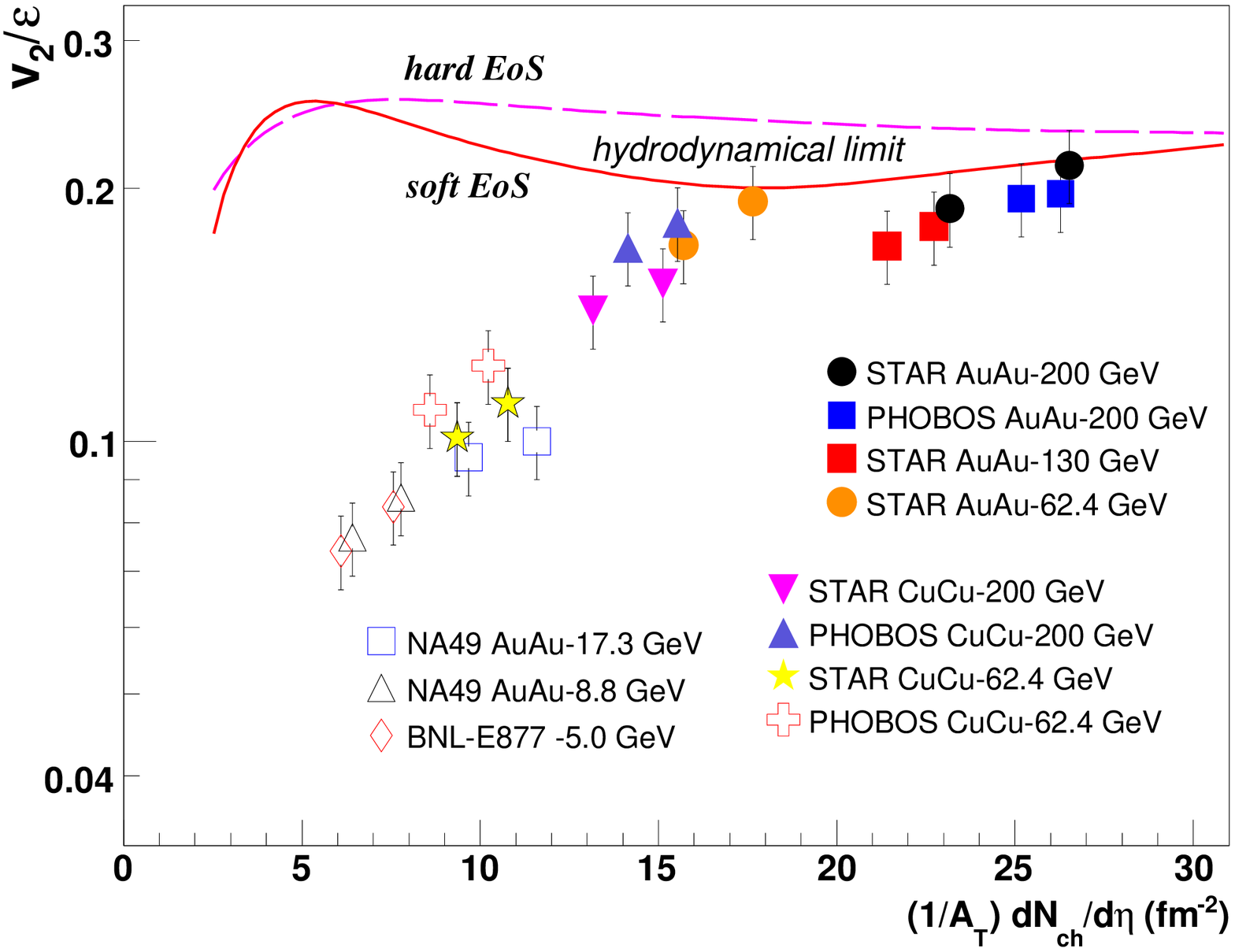}
\vspace{-0.2cm}
\caption{Elliptic flow (normalised by the participant eccentricity~\protect\cite{phobos_qm05}) $v_2/\epsilon$ 
as a function of the hadron rapidity density $\dNdeta$ normalised by the reaction overlap area $A_\perp$, 
measured at SPS~\cite{na49flow} and RHIC~\cite{phobos_qm05,voloshin06}
compared to the ``hydrodynamical limit'' expectations for a fully thermalised system with hard (HRG-like) or 
soft (QGP-like) EoS~\cite{kolb00,teaney_hydro,voloshin_poskanzer}. Adapted from~\protect\cite{voloshin06,phobos_qm05} 
(10\% errors have been added to approx. account for $v_2$, $\epsilon$ syst. uncertainties).}
\label{fig:v2_hydro_limit}
\end{center}
\vspace{-.4cm}
\end{figure} 

\noindent
These results have attracted much attention for several reasons. First, the strong $v_2$ seen in the data is not consistent 
with the much lower values,  $v_2\lesssim$ 6\%, expected by transport models of hadronic matter~\cite{UrQMD} or for 
a partonic system rescattering with perturbative cross sections ($\sigma_{gg}\approx$ 3 mb)~\cite{molnar01}. The 
magnitude, and the $p_T$ and hadron mass dependences of the radial and elliptic  flows below $p_T\approx$ 2 GeV/$c$ 
are, however, well described by {\it ideal} hydrodynamics models whose space-time evolution starts with a 
realistic QGP Equation of State (EoS) with initial energy densities $\varepsilon_0\approx$ 30 GeV/fm$^3$ at thermalization 
times $\tau_0\approx$ 0.6 fm/$c$~\cite{kolb_heinz_rep,huovinen_ruuskanen06,teaney_hydro,hirano} (Fig.~\ref{fig:v2}). 
Second, such a degree of accord between relativistic hydrodynamics and the data was absent at lower CERN-SPS 
energies~\cite{na49flow}. Fig.~\ref{fig:v2_hydro_limit} shows the particle-density dependence of the $v_2$ parameter 
(scaled by the eccentricity of the reaction $\epsilon$ to remove centrality-dependent geometrical effects) measured 
in semicentral collisions at different c.m. energies.
%versus the particle transverse density given by the total charged multiplicity normalised by the overlapping area of the reaction. 
The fact that SPS data lie below the ``hydrodynamical limit'' curves~\cite{voloshin_poskanzer,kolb00} estimated for a 
system completely thermalised, suggests that equilibration is only partially achieved at the top SPS energy~\cite{na49flow}.
RHIC $v_2$ data in the range $\sqrtsnn\approx$ 62 -- 200 GeV~\cite{voloshin06,phnx_v2_62,phobos_qm05} are, however, 
close to the full thermalization expectations. Third, inclusion of viscous (i.e. ``internal dissipation'') corrections to the ideal 
fluid dynamics equations spoils the reproduction of $v_2(p_T)$, especially above $p_T\approx$ 1 GeV/$c$ where even a 
modest viscosity brings $v_2$ towards zero~\cite{teaney03}. Estimates of the maximum amount of viscosity allowed 
by the $v_2(p_T)$ data~\cite{hirano05} give a value for the dimensionless viscosity/entropy ratio close to the 
conjectured universal lower bound,  $\eta/s=\hbar/(4\pi)$, obtained from AdS/CFT calculations~\cite{kovtun04}. 
Similarly, approaches~\cite{teaney_moore04,HvHRapp05} that describe simultaneously the large $v_2(p_T)$ and 
quenching factors $R_{AA}(p_T)$ of heavy-flavour $e^\pm$ require small heavy-quark diffusion coefficients ($3 < 2\pi\,T\,D <6$) 
which correspond to very small shear viscosities, $1 < 4\pi(\eta/s) <2$, and/or very short thermalization times.\\

\noindent
The fast (local) thermalization times supported by the robust collective flow generated in the first instants 
of the reaction, and the good agreement of the $p_T$- and mass-differential spectra and $v_2$ with 
{\it ideal} relativistic hydrodynamics models which assume a fluid evolution with zero viscosity (i.e. with 
negligible internal shear stress), have been presented as evidences that the matter formed at RHIC is  a 
{\it strongly interacting} QGP (sQGP)~\cite{shuryak03,tdlee04,gyulassy_mclerran04,peshier05,heinz05}. 
This new state of matter -- with {\it liquid}-like properties e.g. a Coulomb coupling parameter 
$\Gamma = \mean{E_{pot}}/\mean{E_{kin}} \sim g^2 (4^{1/3}T)/(3T)\sim$ 3 for $g^2 \sim 4-6$ at 
$T\approx$ 200 MeV~\cite{thoma04,peshier05} -- challenges the anticipated paradigm~\cite{shuryak77} 
of a weakly interacting gas of relativistic partons (with $\Gamma \ll$ 1), %\noindent
%All these results put together have justified the assertion that the {\it strongly interacting} QGP (sQGP) 
%produced at RHIC is the most perfect fluid ever observed~\cite{} and, therefore, that the application of 
lending support to the application of strongly-coupled-gauge/weakly-coupled-gravity duality 
techniques~\cite{kovtun04,wiedem06,heavyQ_adscft} to compute relevant sQGP parameters.\\

%Alternatively, it has been proposed~\cite{mueller_visco06} that the system could be instead a weakly-coupled 
%QGP showing large turbulence effects due to Weibel-type instabilities~\cite{strickland06}. Last but not least, 
\noindent
It is worth noting, however, that recent lattice results predict a QCD transition temperature, 
$\Tcrit\approx$ 190 MeV~\cite{latt_Tc_06}, which is $\sim$30--40 MeV larger than the freeze-out temperature 
extracted from observed particle yields in heavy ion experiments (dotted-dashed curve in 
Fig.~\ref{fig:QCD_facets})~\cite{freezeout06,pbm}. Thus, an intermediate regime between the QCD 
transition and freeze-out could exist during which the system created in heavy ion collisions persists in a very 
dense strongly-interacting hadronic phase. %which could be in part responsible of the strongly interacting behaviour observed in the data.
At LHC energies the contribution from the QGP phase to the collective particle flow(s) will be much larger than at RHIC 
or SPS and, therefore, the $v_2$ will be less dependent on the details of the subsequent hadronic phase. The measurement 
of the differential elliptic flow properties in $AA$ collisions at the LHC will be of primary importance to confirm 
or not the sQGP interpretation as well as to search for a possible weakening of the $v_2$ indicative of the existence 
of a weakly interacting QGP phase at higher temperatures than those of the liquid-like state found at RHIC~\cite{thoma04,hirano05}.

%%%%%%%%%%%%%%%%%%%%%%%%%%%%%%%%%%%%%%%%%%%%%%%%%%%%%%%% 

\section{Parton number density and $\qhat$ transport coefficient: High-$p_T$ hadron suppression}
\label{sec:jet_quench}

\begin{figure}[htb]
\centering
\includegraphics[width=8.5cm,height=5.65cm]{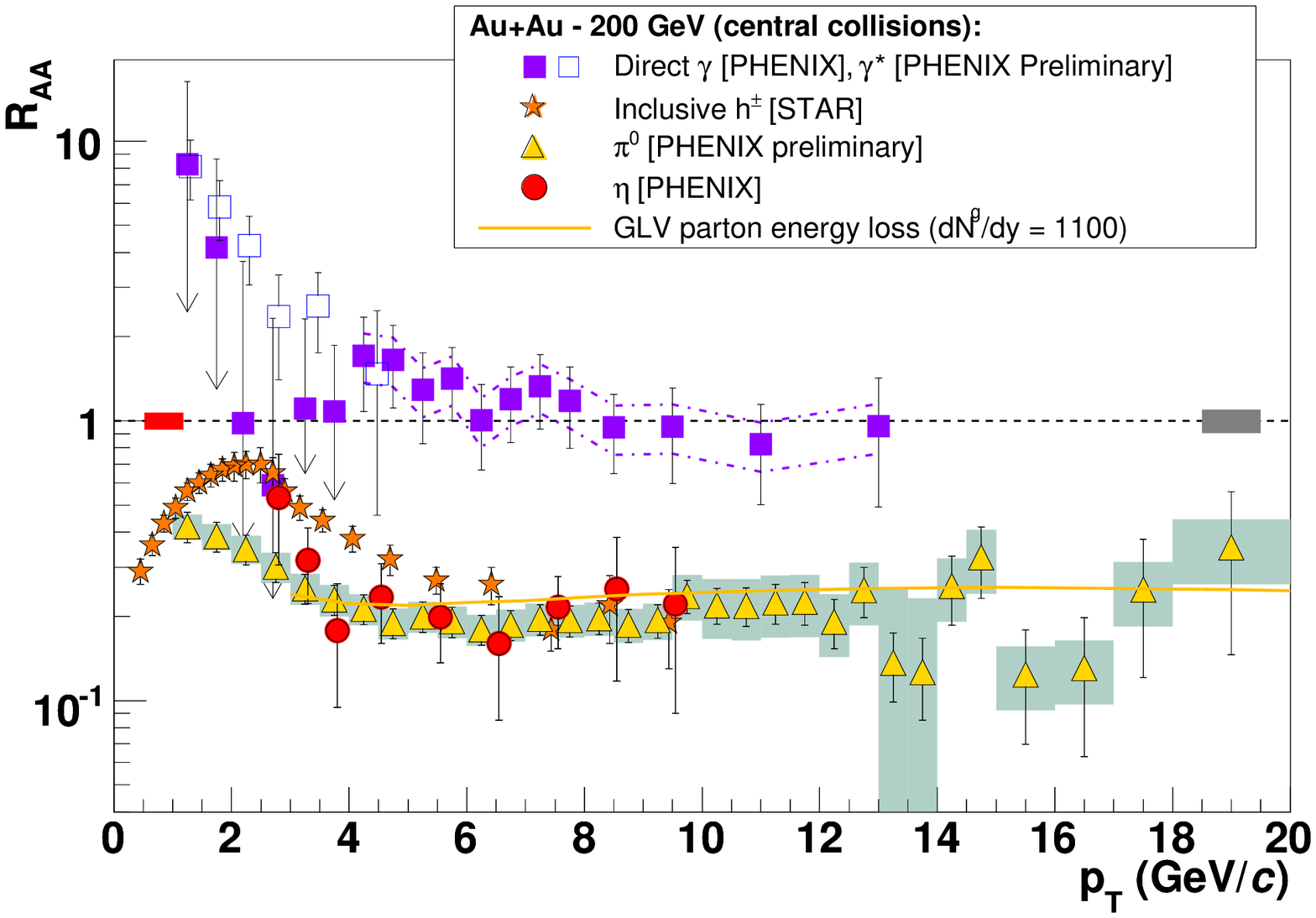}
\includegraphics[width=7.cm,height=5.25cm,clip=true]{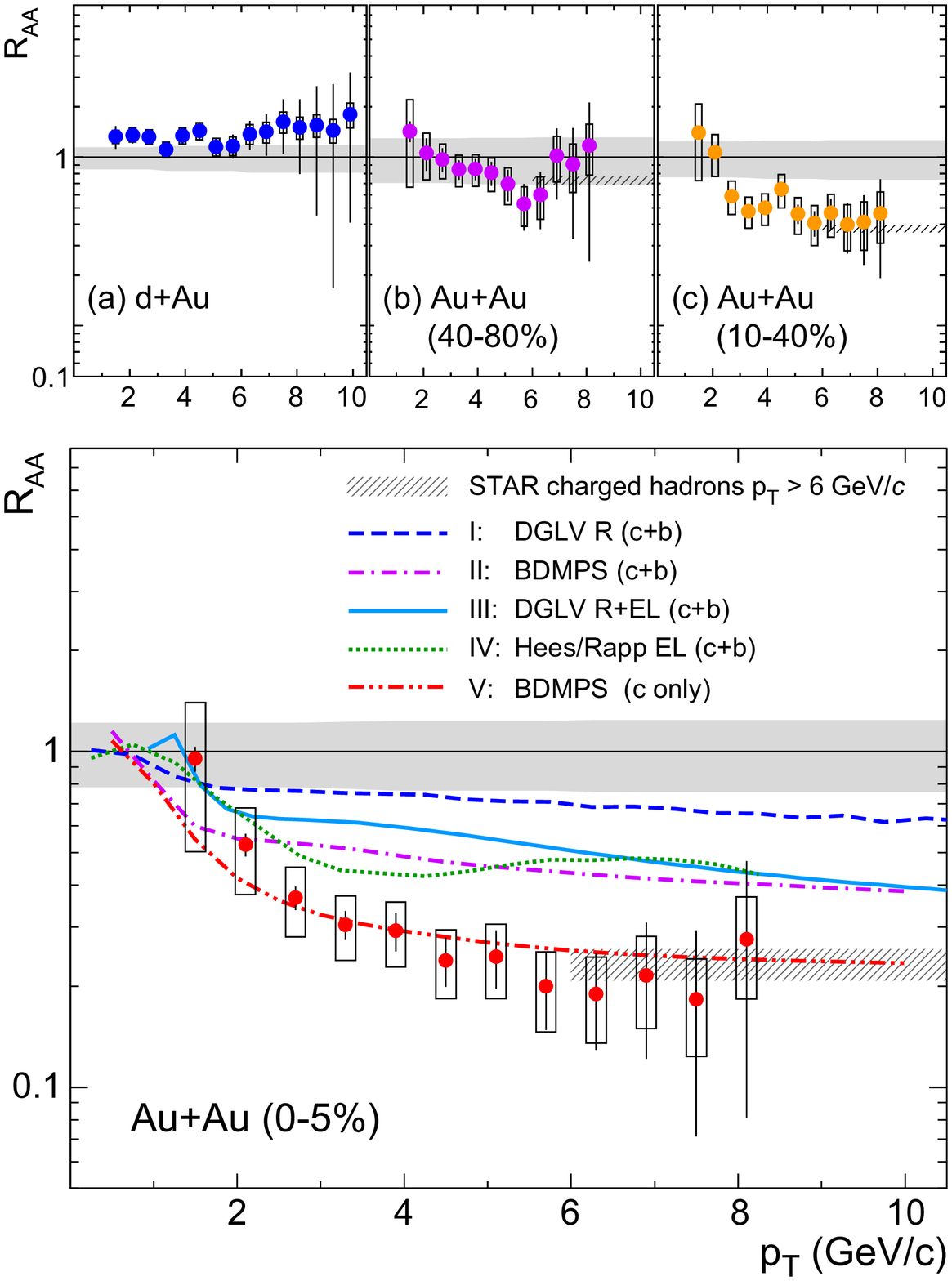}
%\vskip -0.7cm
\caption{Nuclear modification factor $R_{AA}(p_T)$ for $\gamma$, $\pi^{\circ}$, 
$\eta$~\protect\cite{phnx_hipt_pi0_eta_AuAu200}, $\gamma^\star$~\cite{akiba05}, %,phnx_pp_gamma_200}, 
and $h^\pm$~\cite{star_hipt_200} (left) and for ``non-photonic'' $e^\pm$ from D,B 
mesons~\protect\cite{star_nonphoton_elect_AuAupp200} (right) in central $AuAu$ at $\sqrtsnn$ = 200 GeV 
compared to various  parton energy loss model predictions~\cite{vitev_gyulassy,dglv,asw05,HvHRapp05}.}
\label{fig:R_AA_RHIC_200}
\end{figure}

\begin{figure}[htb]
\centering
\includegraphics[width=7.5cm,height=5.75cm]{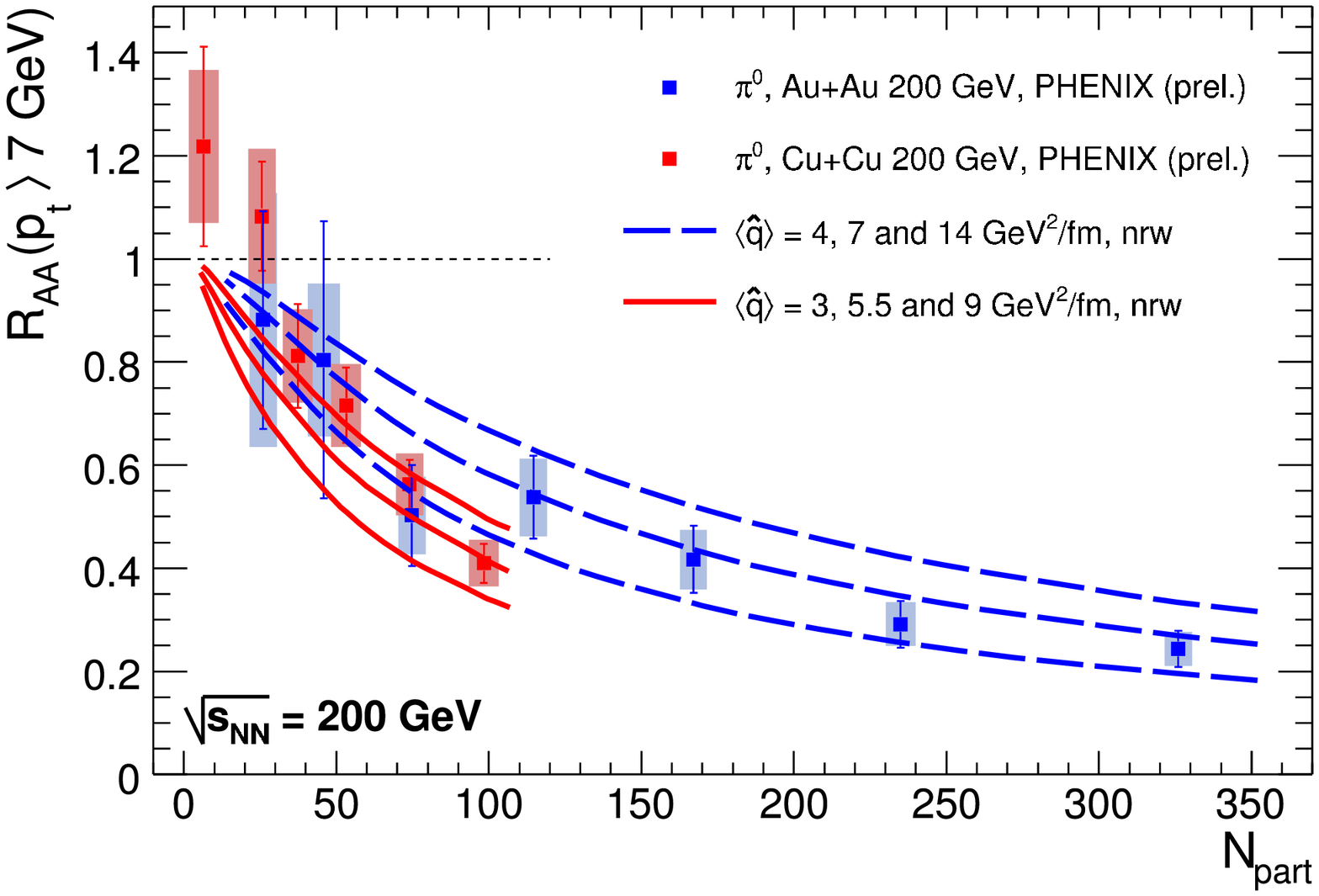}
\includegraphics[width=7.5cm]{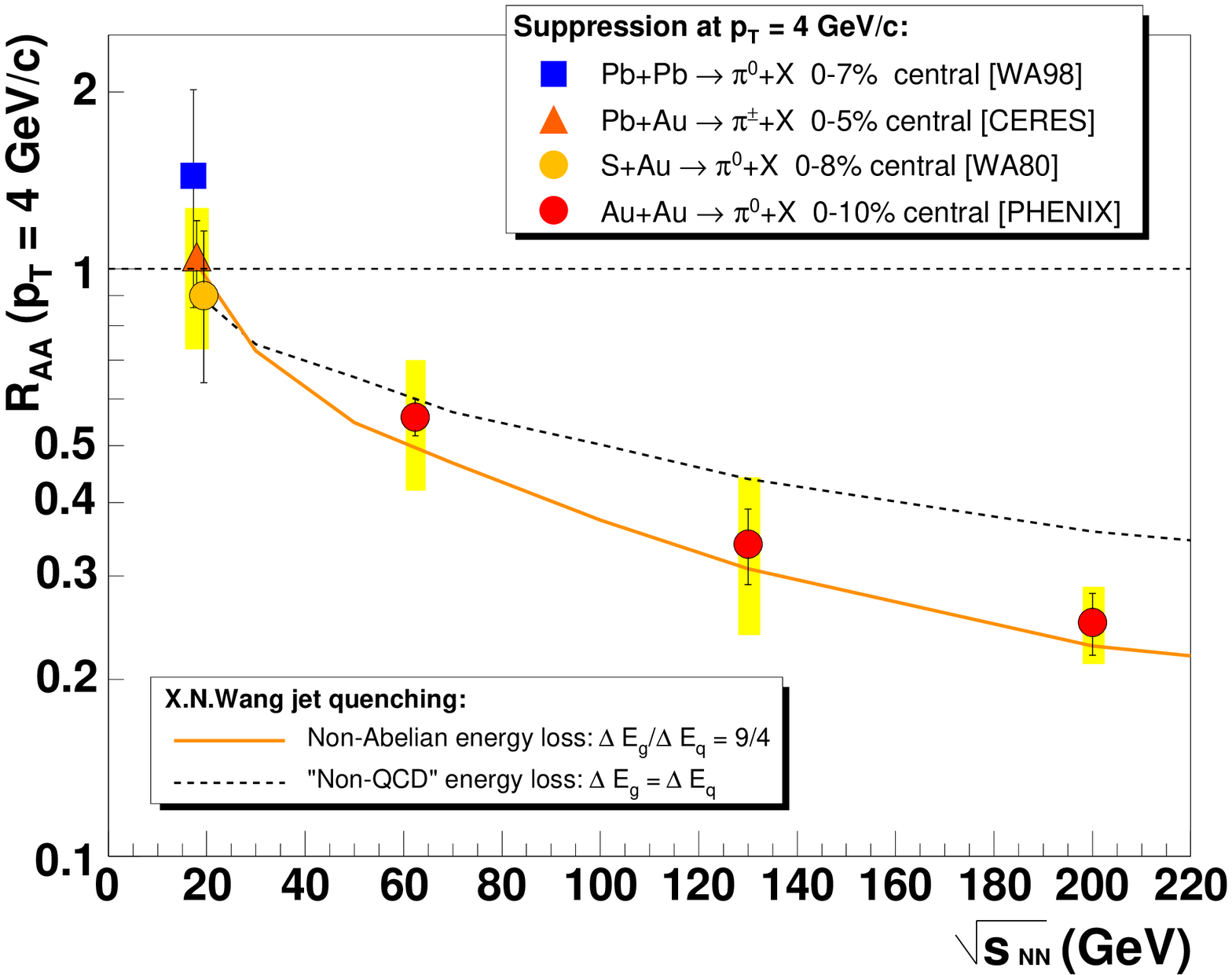}
\vskip -0.cm
\caption{Left: Centrality dependence of $R_{AA}(p_T>$ 7 GeV/$c)$ for $\pi^\circ$ in $AuAu$ and $CuCu$
collisions at $\sqrtsnn$ = 200 GeV compared to the BDMPS-based PQM model~\protect\cite{pqm,loiz06} 
for different values of the $\qhat$ coefficient. Right: Excitation function of the nuclear modification factor, $R_{AA}(\sqrtsnn)$, for 
$\pi^{\circ}$ in central $AA$ reactions at a fixed $p_T$ = 4 GeV/$c$~\protect\cite{dde_hp04}, 
compared to predictions of a jet-quenching model with canonical non-Abelian (solid line) energy loss~\protect\cite{wang05}.}
%and ``non-QCD'' (dashed line) energy losses~\protect\cite{wang05}.}
\label{fig:RAA_data_vs_models}
\end{figure}

Among the most exciting results of the RHIC physics programme is the observed strong suppression of 
high-$p_T$ leading hadron spectra in central $AA$~\cite{rhic_hipt} consistent with the predicted attenuation 
of the parent quark and gluon jets in a dense QCD medium (``jet quenching'')~\cite{bjorken82,gyulassy90}. 
Above $p_T\approx$ 5 GeV/$c$, neutral mesons ($\pi^{\circ}$, $\eta$)~\cite{phnx_hipt_pi0_eta_AuAu200} and 
inclusive charged hadrons~\cite{star_hipt_200} all show a common factor of $\sim$5 suppression compared to an 
incoherent superposition of $pp$ collisions (Fig.~\ref{fig:R_AA_RHIC_200}, left). Such a  significant suppression 
was not observed at SPS where, after reevaluating the $pp$ baseline spectrum~\cite{dde_hipt_sps,wa98_pi0}, 
the central-$AA$ meson spectra show a $R_{AA}$ around unity (Fig.~\ref{fig:RAA_data_vs_models}, right) 
probably due to the cancellation of a $\sim$50\% final-state suppression by initial-state Cronin 
broadening~\cite{dde_hipt_sps,antinori05,blume06}. At RHIC, the $R_{AA}\approx$ 1 perturbative expectation 
which holds for other hard probes such as ``colour blind''  photons~\cite{phnx_gamma_AuAu200} -- and for 
high-$p_T$ hadrons in $dAu$ reactions (Fig.~\ref{fig:pp_rhic_vs_nlo}, right) -- is badly broken 
($R_{AA}\approx$ 0.2) in central $AuAu$ collisions. This strongly supports the picture of partonic 
energy loss in {\it final-state} interactions within the dense  matter produced in the reaction.\\

\noindent
The dominant contribution to the energy loss is believed to be of non-Abelian radiative nature 
(i.e. due to gluon radiation) as described in the GLV~\cite{glv,vitev_gyulassy} 
and BDMPS~\cite{bdmps,wiedemann} (or LPCI~\cite{lpci})
formalisms. In the GLV approach, the initial gluon density $dN^g/dy$ of the expanding plasma 
(with transverse area $A_\perp$ and length $L$) can be estimated from the measured energy loss 
$\Delta E$:
\begin{equation}
\Delta E \propto \alpha_S^3\,C_R\,\frac{1}{A_\perp}\frac{dN^g}{dy}\,L\mbox{ .}
\label{eq:glv}
\end{equation}
where $C_R$ is the Casimir colour factor of the parton (4/3 for quarks, 3 for gluons).
In the BDMPS framework, the transport coefficient\footnote{Technically, the $\hat{q}$ parameter can be 
identified with the coefficient in the exponential of an adjoint Wilson loop averaged over the medium length: 
$\mean{W^A(C)}\equiv exp\left[(-1/4\sqrt{2})\hat{q}L^-L^2\right]$~\cite{wiedem06}.}  
$\langle\hat{q}\rangle$, characterizing the squared average momentum transfer of the hard parton 
per unit distance, can be derived from the average energy loss according to:
\begin{equation}
\langle\Delta E\rangle \propto \alpha_S\,C_R\,\langle\hat{q}\rangle\,L^2.
\label{eq:bdmps}
\end{equation}
\noindent
From the general Eqs. (\ref{eq:glv}) and (\ref{eq:bdmps}), very large initial gluon rapidity densities, 
$dN^g/dy\approx$ 1100 $\pm$ 300~\cite{vitev_gyulassy}, or equivalently, transport coefficients 
$\qhat\approx$ 11 $\pm$ 3 GeV$^2$/fm~\cite{pqm,eskola04,asw05,loiz06}, are required in order to 
explain the observed amount of hadron suppression at RHIC. The corresponding values for SPS are
$dN^g/dy\approx$ 400 $\pm$ 100 and $\qhat\approx$  3.5 $\pm$ 1 GeV$^2$/fm~\cite{dde_hp04}. 

\noindent
Most of the empirical properties of the quenching factor for light-flavour hadrons -- magnitude, $p_T$-, centrality-,
$\sqrtsnn$- dependences of the suppression -- are in quantitative agreement with the predictions of non-Abelian 
parton energy loss models (Fig.~\ref{fig:RAA_data_vs_models}). However, the fact that the high-$p_T$  
$e^\pm$ from semi-leptonic $D$ and $B$ decays is as suppressed as the light hadrons in central $AuAu$ 
(Fig.~\ref{fig:R_AA_RHIC_200}, right)~\cite{phnx_hipt_nonphoton_elect_AuAu200,star_nonphoton_elect_AuAupp200} 
is in apparent conflict with the robust $\Delta E_{\rm heavy-Q} < \Delta E_{\rm light-q} <  \Delta E_{g}$ prediction of radiative 
energy loss models. Since the gluonsstrahlung probability of quarks and gluons is completely determined by 
the gauge structure (Casimir factors) of SU(3), the colour octet gluons (which fragment predominantly into
{\it light} hadrons) are expected to lose energy at $C_A/C_F$ = 9/4 times the rate of quarks. In addition,
massive $c, b$ quarks are expected to lose less energy than light ones due to their suppressed small-angle 
gluon radiation already in the vacuum (``dead-cone'' effect)~\cite{doksh_kharzeev}.
%The probability for a gluon (quark) to radiate a 
%gluon is proportional to the color factor $C_A$ = 3 ($C_F$ = 4/3). In the asymptotic limit, 
%in which the radiated gluons carry a small fraction of the original parton momentum, 
%and neglecting the splitting of gluons to quark-antiquark pairs (proportional 
%to the smaller color factor $T_R$ = 1/2), the average number of gluons radiated by 
%a gluon is, therefore, a factor $C_A/C_F$ = 9/4 higher than the number of gluons radiated 
%by a quark~\cite{ellis96}.}
In order to reproduce the high-$p_T$ open charm/bottom suppression, jet quenching models require either 
initial gluon rapidity densities ($dN^g/dy\approx$ 3000)~\cite{dglv} inconsistent with the total hadron 
multiplicities, $dN^g/dy\approx 1.8\,\dNdeta$~\cite{dde_hp04} or with the $dN^g/dy$ needed 
to describe the quenched light hadron spectra, or they need a smaller relative contribution of $B$ relative to 
$D$ mesons than theoretically expected in the measured decay electron $p_T$ range~\cite{asw05}. 
This discrepancy\footnote{Note, however, that the theoretical and experimental control of the 
$pp\rightarrow D,B+X$ reference (Fig.~\ref{fig:pp_rhic_vs_nlo}, left) is not as good as for the light 
hadron spectra~\cite{asw05}.} may point to an additional contribution from elastic (i.e. non-radiative) 
energy loss~\cite{mustafa03,peshier06} for heavy-quarks~\cite{dglv} which was considered negligible 
so far~\cite{gyulassy90}. The unique possibility at the LHC to fully reconstruct jets~\cite{yr_lhc_jets}, 
to tag them with prompt $\gamma$~\cite{jet_gamma_tagg} or $Z$~\cite{jet_Z_tagg}
%-- %free of the ``surface'' bias of leading hadrons allowing to better constrain the $\qhat$ values -- 
and to carry out detailed studies in the $c,\,b$ quark sector~\cite{yr_lhc_heavyQ,dainese_hp06} will be 
very valuable to clarify the response of strongly interacting matter to fast heavy-quarks, and will provide 
accurate information on the transport properties of QCD matter~\cite{wiedem06,heavyQ_adscft}.

%%%%%%%%%%%%%%%%%%%%%%%%%%%%%%%%%%%%%%%%%%%%%%%%%%%%%%%%

\section{Propagation of collective perturbations in QCD matter: Distorted di-jet correlations}
%\section{Excitation of collective modes (speed of sound $c_s$): Distorted di-jet correlations}
%\section{Modified semihard di-hadron $\phi$ correlations: QGP speed of sound $c_s$ ?}
\label{sec:dijets}

\begin{figure}[htb]
\centering
\includegraphics[width=7.6cm,height=6.8cm]{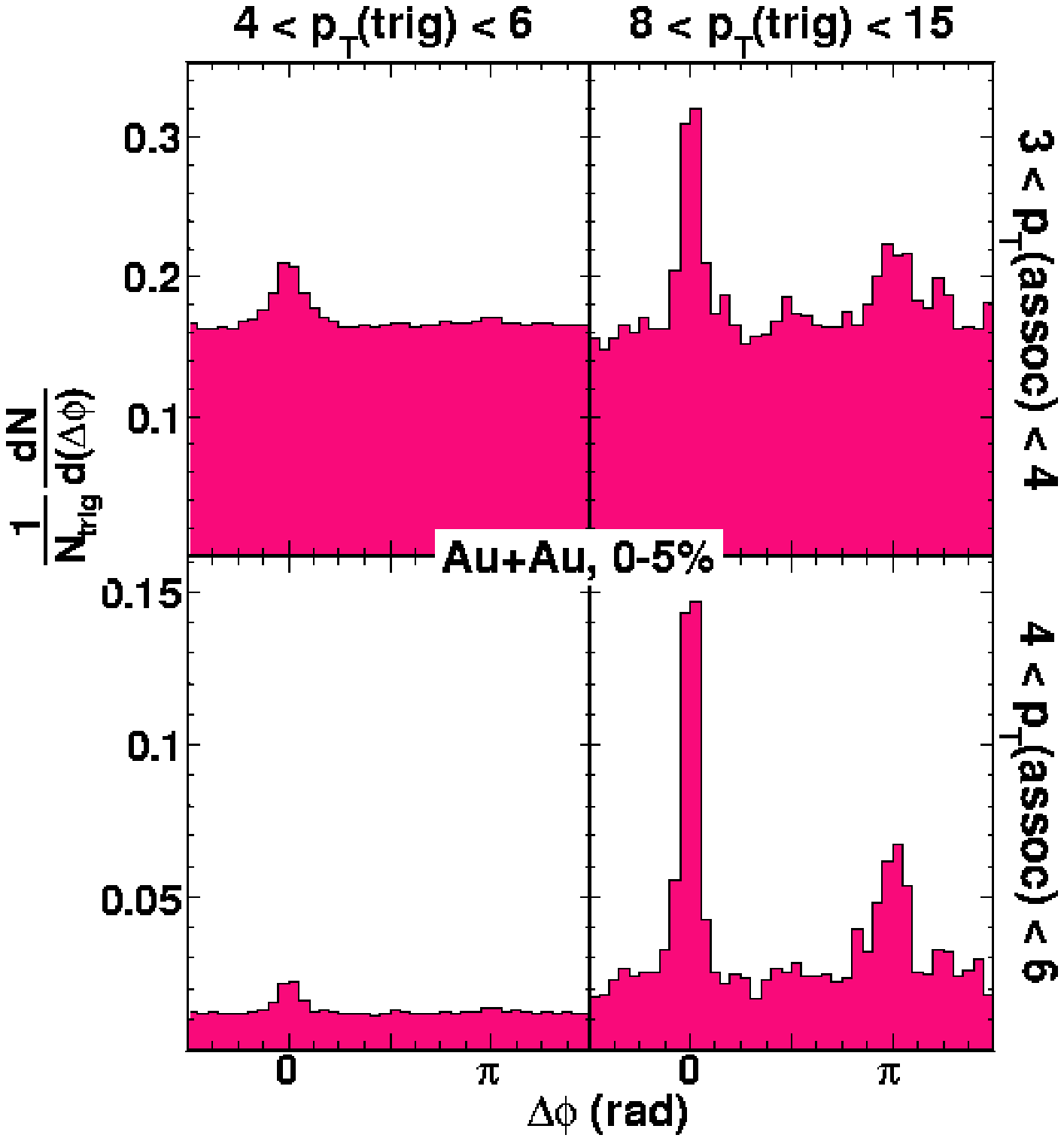}
\includegraphics[width=7.6cm,height=6.75cm]{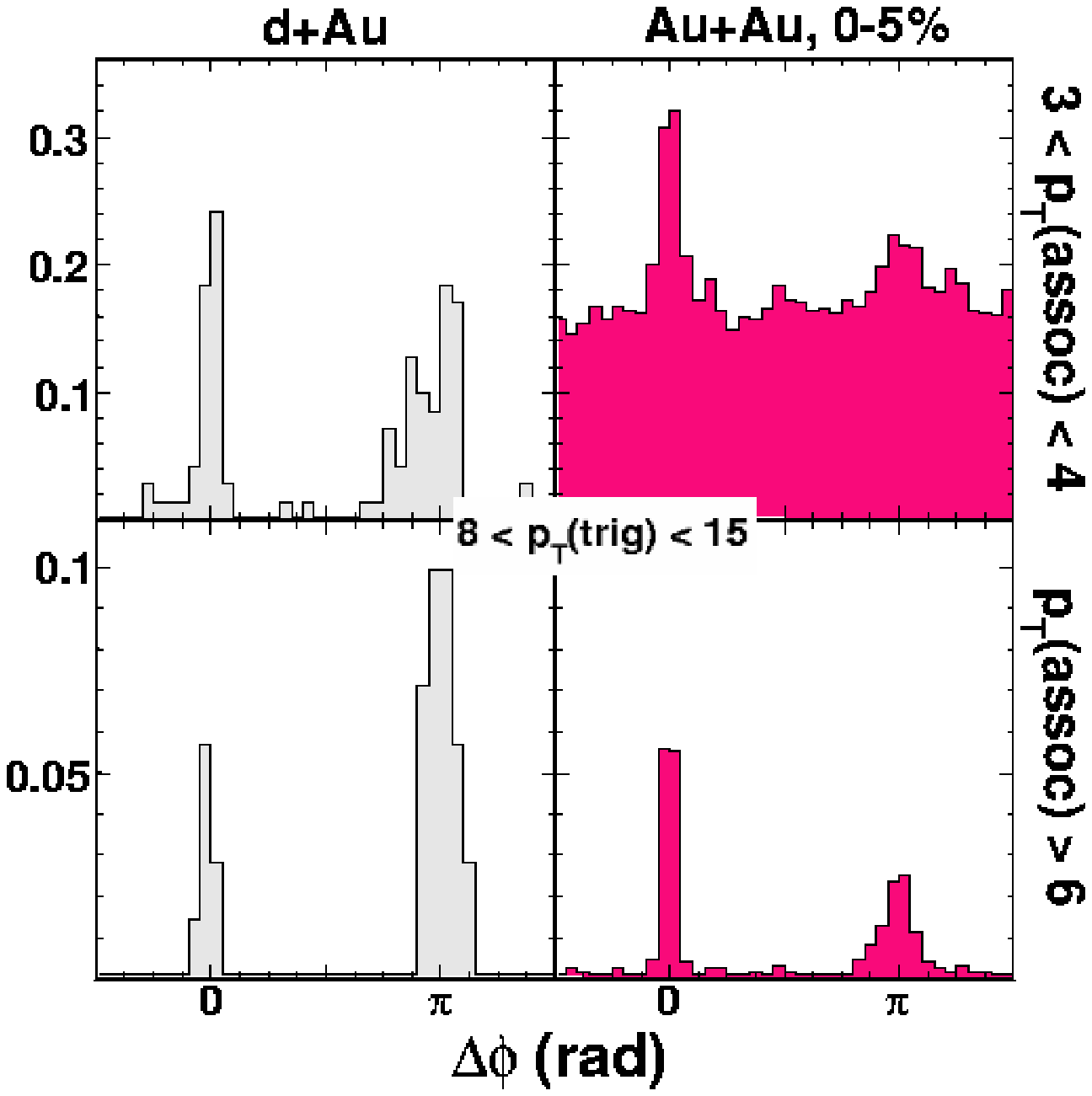}
%\vskip -0.cm
\caption{Angular correlations of high-$p_T$ charged hadron pairs measured in: 0-5\% central $AuAu$ events
in various $p_{T,\,trig}$ and $p_{T,\,assoc}$ (GeV/$c$) ranges (left), and in $dAu$ and 0-5\%-central $AuAu$ for
fixed  $p_{T,\,trig}$ = 8 -- 15 GeV/$c$ and two $p_{T,\,assoc}$ ranges~\protect\cite{star_jet_punchthrough} (right).}
\label{figs:star_dijets}
\end{figure}

Full jet reconstruction in $AA$ collisions with standard jet algorithms~\cite{jet_algo} is unpractical 
at RHIC energies due to low cross-sections for high-$E_T$ jets and the overwhelming background 
of soft particles in the underlying event (only above $\sim$30 GeV are jets above the background). 
Instead, jet-like correlations at RHIC are conventionally measured on a statistical basis by selecting high-$p_T$ 
{\it trigger} particles and measuring the azimuthal ($\Delta\phi = \phi - \phi_{trig}$) and rapidity 
($\Delta\eta = \eta - \eta_{trig}$) distributions of {\it associated} hadrons ($p_{T,\,assoc}<p_{T,trig}$) 
relative to the trigger:
\begin{equation}
C(\Delta\phi,\Delta\eta) = \frac{1}{N_{trig}}\frac{d^2N_{pair}}{d\Delta\phi d\Delta\eta}.
\end{equation}
Combinatorial background contributions, corrections for finite pair acceptance, and the superimposed 
effects of global azimuthal modulations (elliptic flow) are taken into account with different 
techniques~\cite{star_hipt_awayside,star_hipt_etaphi,phnx_machcone,nuchem05}.
In $pp$ or $dAu$ collisions, a dijet signal appears as two distinct back-to-back Gaussian-like peaks around 
$\Delta\phi$ =  0 (near-side) and $\Delta\phi=\pi$ (away-side) ($dAu$ panel in Fig.~\ref{figs:star_dijets}). 
At variance with this standard dijet topology in the QCD vacuum, early STAR results for semihard jets in 
central $AuAu$ reactions~\cite{star_hipt_awayside} showed a complete disappearance of the opposite 
side peak for 3 $ < p_{T,\,assoc}< 4 <p_{T,\,trig}<$ 6 GeV/$c$ while the near-side correlation 
remained unchanged (Fig.~\ref{figs:star_dijets}, leftmost panel). Such a monojet-like 
topology confirmed a jet-quenching picture where a $2\rightarrow 2$ hard scattering takes place near the 
surface of the system with the trigger parton being unaffected and the away-side parton losing energy while 
traversing a medium opaque to coloured probes. For rising $p_{T,\,trig}$, the away-side parton is seen to 
increasingly ``punchthrough'' the medium~\cite{star_hipt_awayside}, although the azimuthally-opposite 
correlation strength is still significantly reduced compared to $dAu$ (Fig.~\ref{figs:star_dijets}).

\noindent
The estimated energy loss, Eqs.~(\ref{eq:glv}) and (\ref{eq:bdmps}), of the quenched partons is very large, 
up to $\Delta E_{\mbox{\tiny{\it loss}}}\approx$ 3 GeV/fm for a 10 GeV parton~\cite{loiz06} and 
most of the (mini)jets, apart from those close to the surface, dump a significant fraction of their energy and 
momentum in a cell of about 1 fm$^3$ in the rest frame of the medium. Since energy and momentum are conserved, 
the fragments of the quenched parton are either shifted to lower energy ($p_T<$2 GeV/$c$) and/or scattered 
into a broadened angular distribution. Both softening and broadening of the away-side distribution are seen in the 
data~\cite{star_hipt_etaphi} when the $p_T$ threshold of the away-side associated hadrons is {\it lowered}. 
\begin{figure}[htbp]
\includegraphics[width=7.5cm,height=7.cm]{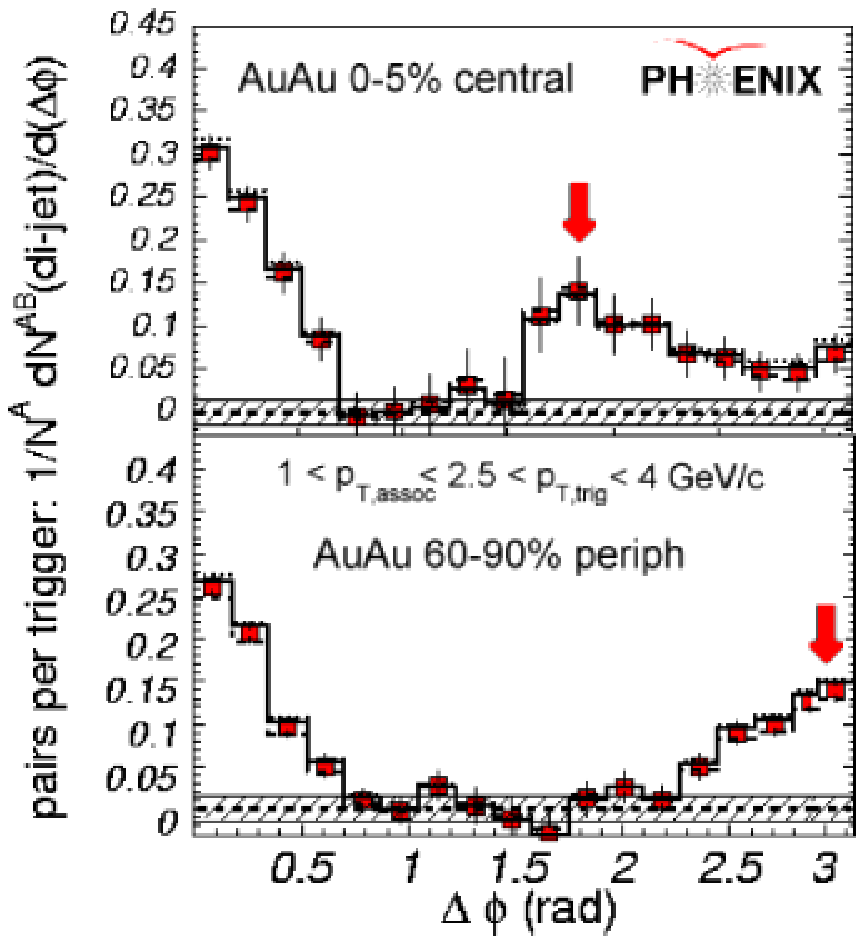}%}
%\includegraphics{denterria_phenix_machcone.eps}%}
%\hskip 0.5cm
\includegraphics[width=8.4cm,height=5.5cm]{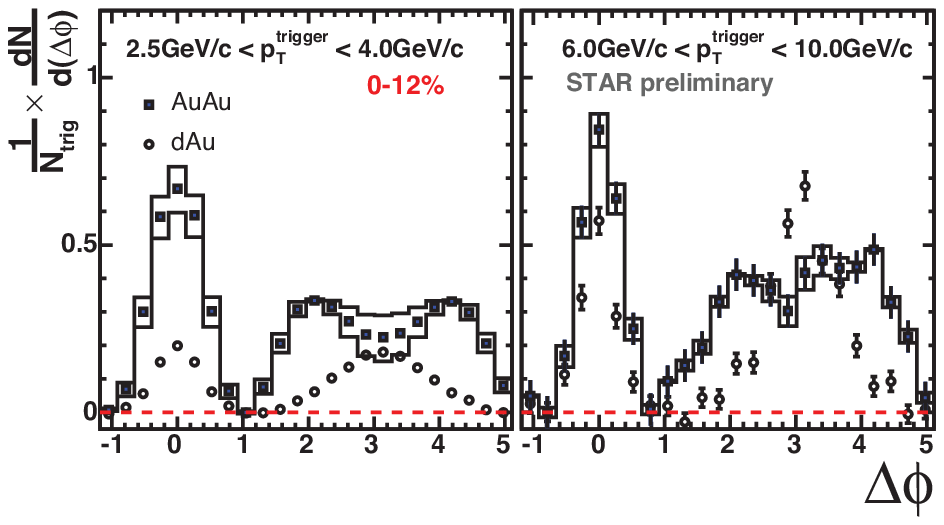}
%\includegraphics{denterria_dNdphi_star_hp06b.eps}
%\vskip -0.7cm
\caption{Azimuthal distributions of semihard hadrons ($p_{T,\,assoc}$ = 1 -- 2.5 GeV/$c$) 
relative to a higher $p_T$ trigger hadron measured at RHIC. 
Left: PHENIX data in central (top) and peripheral (bottom) $AuAu$~\protect\cite{phnx_machcone}
(the arrows indicate the local maxima in the away-side hemisphere). 
Right: STAR data in central $AuAu$ (squares) and $dAu$ (circles) for two ranges of $p_{T,\,trig}$~\protect\cite{star_machcone}.}
\label{fig:dNdphi}
\end{figure}
Fig.~\ref{fig:dNdphi} shows the dihadron azimuthal correlations $dN_{pair}/d\Delta\phi$ in central $AuAu$ collisions 
for $p_{T,\,assoc}$ = 1 -- 2.5 GeV/$c$~\cite{phnx_machcone,star_machcone}. In this semi-hard range, the away-side 
hemisphere shows a very unconventional angular distribution with a ``dip'' at $\Delta\phi\approx\pi$ and two 
neighbouring local maxima at $\Delta\phi\approx\pi\pm$ 1.1. Such a non-Gaussian ``volcano''-like profile has been 
explained as due to the preferential emission of energy from the quenched parton at a finite angle with respect to the 
jet axis. This could happen in a purely radiative energy loss scenario~\cite{salgado06} but more intriguing 
explanations for the conical-like pattern have been put forward based on the dissipation of the lost energy into a collective 
mode of the medium which generates a wake of lower energy gluons with Mach-~\cite{mach1,mach2,rupp05} or 
\v{C}erenkov-like~\cite{rupp05,cerenkov1,cerenkov2} angular emissions. 
In the Mach-cone scenario, the {\it speed of sound}\footnote{The speed of sound %an ultrarelativistic fluid 
is a simple proportionality constant relating the fluid pressure and energy density: $P=c_s^2\;\varepsilon$.}
of the traversed matter, $c_s^2 = \partial P/\partial\varepsilon$, can be determined from the characteristic supersonic 
angle of the emitted secondaries: %with respect to the (quenched) jet axis
$\cos(\theta_{M}) = \mean{c_s}$, where $\theta_{M}$ is the Mach shock wave angle and $\mean{c_s}$ 
the time-averaged value of the speed of sound of the medium traversed by the parton.
%In the ``Mach cone'' scenario~\cite{mach,rupp05}, the local maxima (red arrows in the plots) 
%in central $AuAu$ at an angle $\Delta\phi \approx \pi\pm 1.2$ relative to the high-$p_T$ 
%trigger are caused by the Mach shock of the supersonic recoiling (quenched) parton through the medium. 
The resulting preferential emission of secondary partons from the plasma at a {\it fixed} 
angle $\theta_{M} \approx 1.1$, %relative to the away-side quenched parton. 
%angle $\theta_{M} \approx \arccos(\bar{c}_{s})\approx 1.2$  %relative to the away-side quenched parton. 
%caused by its supersonic propagation of the plasma particles traveling at an angle with respect to the jet axis:
%yields (Eq.~\ref{eq:mach}) a time-averaged value of the speed sound in the expanding medium, 
%$\bar{c}_{s}\approx$ 0.33 inbetween the QGP, $c_s = 1/\sqrt{3}$, and hadron resonance gas, $c_s = \sqrt{0.2}$, values.
yields a value $\mean{c_{s}}\approx$ 0.45 which is larger than that of a hadron-resonance gas 
($c_s \approx$ 0.35)~\cite{alam03}, and not far from that of a deconfined QGP\footnote{Note that although
lattice calculations indicate that there are $\sim$30\% deviations from the ideal-gas limit in the 
$s(T)$,$P(T)$ and $\varepsilon(T)$ dependences up to very high $T$'s~\cite{latt}, the ideal-gas relation 
$\varepsilon \approx 3 P$ (as well as other ratios of thermodynamical potentials) approximately holds 
above $\sim 2\,\Tcrit$.}
% even for regions well below the ideal gas limit %. Indeed $\varepsilon \approx 3 P$ holds already for temperatures 
%slightly above $T_c$.} 
($c_s = 1/\sqrt{3}$). Experimental confirmation of the Mach-cone picture or other alternative emission mechanisms
for the associated particles in the quenched jet requires more detailed differential studies~\cite{roy06} such as the 
ongoing analysis of three- and many-particle azimuthal correlations~\cite{leeuwen06,ajitHP06}.

%%%%%%%%%%%%%%%%%%%%%%%%%%%%%%%%%%%%%%%%%%%%%%%%%%%%%%%%

\section{Bulk hadronization: Enhanced baryon yields/flows at intermediate $p_T$}
%\section{Bulk hadronization: breaking of FF universality at intermediate $p_T$}
%\section{Hadronization via recombination: enhanced baryon yields/flows at intermediate $p_T$}
\label{sec:recomb}

\begin{figure}[htb]
\centering
\includegraphics[width=8.6cm,height=6.cm]{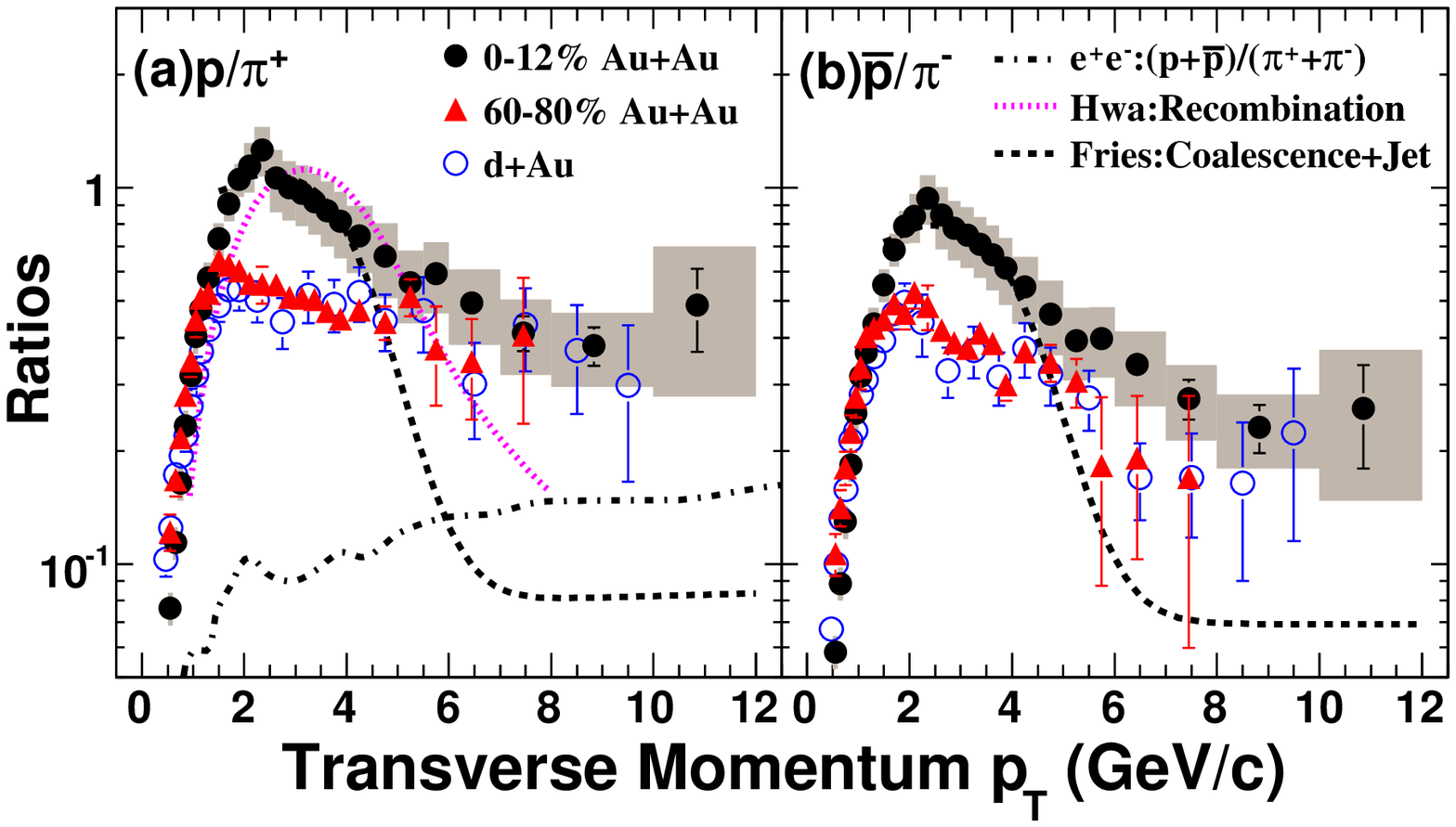}
\hspace{-0.8cm}
\includegraphics[width=7.6cm,height=5.5cm]{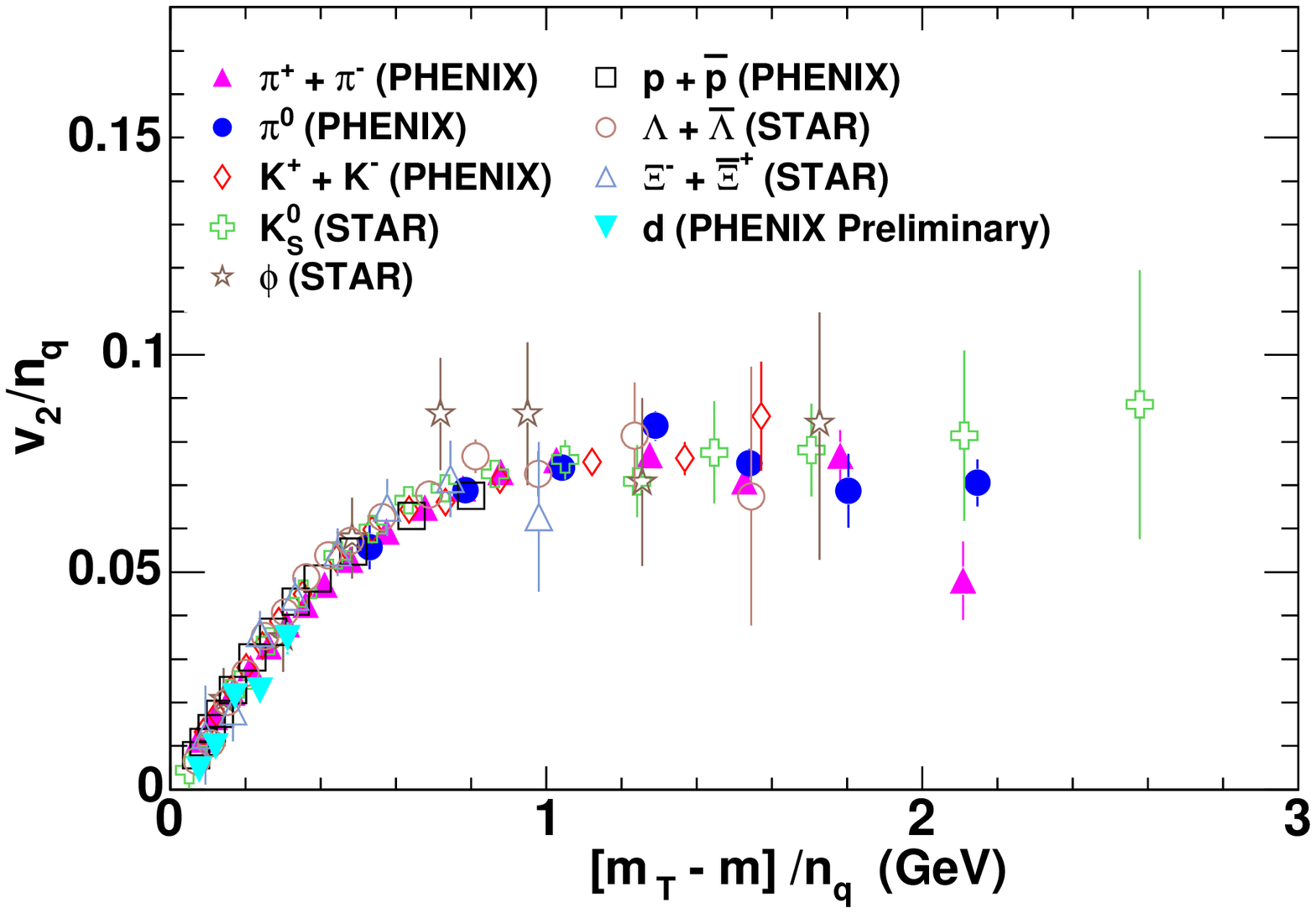}
\vskip -0.cm
\caption{Left: Proton/pion in $dAu$ and $AuAu$ collisions at $\sqrtsnn$ = 200 GeV~\cite{star_hipt_baryons}
compared to the ratio in light quark fragmentation in $e^+e^-$ at $\sqrts$ = 91.2 GeV (dotted-dashed line)
and to parton coalescence predictions~\cite{reco}. Right: Elliptic flow parameter $v_2$ for all  
hadrons at RHIC normalised by the number of constituent quarks of each species ($n_q$ = 2,3 for mesons,baryons) 
$vs$ the transverse kinetic energy $KE_T=m_T-m$ normalised also by $n_q$~\cite{phnx_ppg062,issah06}.}
\label{fig:baryons}
\end{figure}

The increasingly suppressed production of mesons above $p_T\approx$ 2 GeV/$c$ in central $AuAu$ 
reactions at RHIC contrasts with the simultaneous {\it unsuppressed} %baryon 
$p,\bar{p}$~\cite{phnx_hipt_ppbar,phnx_hipt_200,star_hipt_baryons} and 
$\Lambda,\bar{\Lambda}$~\cite{star_hipt_lambdas} yields in the range 
$p_T\approx$ 2 -- 4 GeV/$c$. The intermediate-$p_T$ range in $AuAu$ reactions features an 
anomalous baryon/meson$\sim$ 0.8 ratio which is roughly four times higher than 
in more elementary $pp$ or $e^+e^-$ interactions (Fig.~\ref{fig:baryons}, left). 
Semihard (anti)protons show an enhancement with respect to the ``$x_T$ scaling'' expectation 
for the ratio of perturbative cross sections at different c.m. energies~\cite{mjt}, whereas 
$x_T$ scaling holds for {\it all} hadrons (mesons) measured in $pp$ ($AuAu$) collisions at 
RHIC~\cite{phnx_hipt_200,mjt,star_pp_200GeV_xt_scaling}. Only above 
$p_T\approx$ 6 GeV/$c$~\cite{star_hipt_baryons} (Fig.~\ref{fig:baryons}, left) 
the meson/baryon ratio is again consistent with the expected yields obtained from universal 
fragmentation functions. Not only their spectra are enhanced, but at $p_T$ = 2 GeV/$c$ the $v_2$ 
elliptic flow parameter of baryons exceeds that of mesons ($v_2^{meson}\approx$~0.16, 
see Fig.~\ref{fig:v2}, right) and keeps increasing up to $p_T\sim$ 4 GeV/$c$ when it finally 
saturates at $v_2^{\tiny{\ensuremath{\it baryons}}}\approx$~0.22~\cite{phnx_v2_id,star_v2_id}.
All those observations clearly indicate that standard hadron production via (mini)jet fragmentation is 
not sufficient to explain the RHIC data for baryons at transverse momenta of a few GeV/$c$.
\vspace{-0.5cm}

%Baryons and mesons v2 saturate at significantly different values of pT with baryons having an
%anisotropic flow larger than the mesons for $p_T\sim$2 GeV/$c$
\noindent
The observed baryon nuclear modification factors close to unity, the high baryon/meson ratios, 
and their large elliptic flow have lent support to the existence of an extra mechanism 
of baryon production in $AuAu$ based on quark coalescence in a dense partonic medium~\cite{reco}.
Recombination models compute the spectrum of hadrons as a convolution of Wigner functions with 
single parton thermal distributions leading to an exponential distribution which dominates 
over the standard power-law fragmentation regime below $p_T\sim$ 6 GeV/$c$. Coalescence models 
can thus reproduce the enhanced baryon production and predict also that the elliptic flow of any 
hadron species should follow the underlying partonic flow scaled by the number $n$ of (recombined) 
constituent quarks in the hadron: $v_2(p_T)=n\,v_2^{q} (p_T/n)$, $n=2,3$ 
for mesons and baryons, respectively~\cite{molnar_voloshin}. Such a quark-scaling law 
for the elliptic flow is confirmed by the data~\cite{phnx_v2_id,star_v2_id}. 
Figure~\ref{fig:baryons} right, shows a recent variation of the quark-number scaling law for $v_2$
that uses the transverse kinetic energy ($KE_T=m_T-m$, $m_{T} = (p_T^2+m^2)^{1/2}$) rather 
than the $p_T$ and seems to account perfectly for the scaled $v_2$ of all measured species also in 
the soft hydrodynamical regime below $p_T\sim$2 GeV/$c$~\cite{phnx_ppg062,issah06}. 
The overall success of valence quark coalescence models to explain hadron production in the semi-hard 
regime highlights the role of {\it thermalised} degrees of freedom in the produced system with 
{\it partonic} (as opposed to hadronic) quantum numbers.

%%%%%%%%%%%%%%%%%%%%%%%%%%%%%%%%%%%%%%%%%%%%%%%%%%%%%%%%

\section{Temperature and equation-of-state (EoS): Thermal photons}
\label{sec:photons}

In order to describe the transient systems produced in $AA$ collisions in terms of {\it thermodynamical}
variables ($T$, $\varepsilon$, $s$, etc.) linked by an EoS which can be compared to lattice QCD expectations, 
it is a prerequisite to establish that the underlying degrees of freedom form, at some stage of the reaction, 
a statistical ensemble. Proving that {\it local}\footnote{Note that, by simple causality arguments, 
{\it global} equilibrium in a finite system with radius $R_{A}\approx$ 7 fm can only occur for 
time-scales $\tau\gtrsim$ 7 fm/$c$.} thermalization has been attained in the course of the collision is 
thus a crucial issue both experimentally and theoretically~\cite{baier00}. The large  elliptic flow signal 
observed in the data strongly supports the idea of fast thermalization as discussed in Section~\ref{sec:sQGP}.
The identification of real and/or virtual $\gamma$ radiation from the produced ``fireball'' with properties
consistent with a thermal distribution would in addition allow us to determine the underlying temperature 
and EoS of the system.  Hydrodynamical~\cite{huovinen_ruuskanen06} and parton transport~\cite{bass} 
calculations indicate that the same cascade of secondary parton-parton collisions that drives the system 
towards equilibration in the first tenths of fm/$c$ results in an identifiable emission of thermal radiation 
above the prompt perturbative yield in $AA$ reactions in a window $p_T\approx$ 1 -- 3 GeV/$c$.

\begin{figure}[!htb]
\centering
\includegraphics[width=10cm,height=6.8cm]{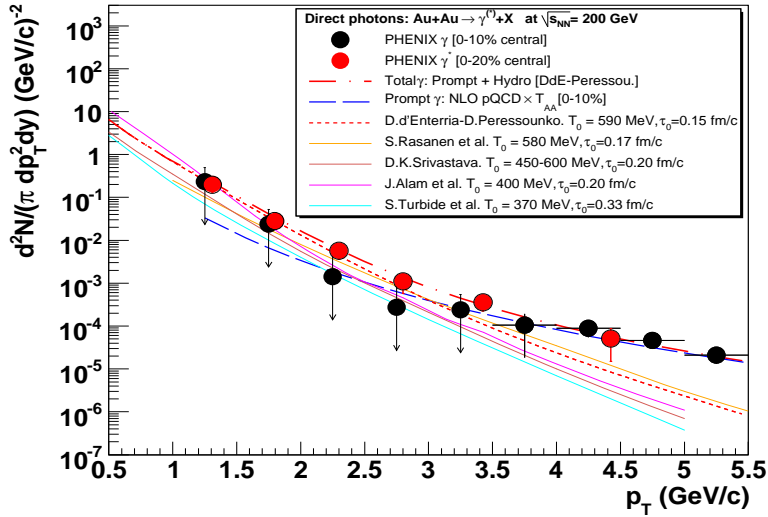}
\vskip -0.3cm
\caption{Direct $\gamma,\gamma^\star$ spectra measured by PHENIX~\cite{phnx_gamma_AuAu200,akiba05} in central 
$AuAu$ collisions at $\sqrtsnn$ = 200 GeV compared to NLO pQCD~\cite{vogelsang_gamma} and 
different hydrodynamics predictions~\cite{dde_peressou} including a QGP phase with $T_0$ = 350-600 MeV.}
\label{fig:thermal_photons}
\end{figure}

\noindent
The preliminary $\gamma$ spectrum measured in central $AuAu$ by PHENIX~\cite{akiba05,bathe05} is 
consistent with the sum of such perturbative (prompt) plus thermal (secondary) contributions 
(Fig.~\ref{fig:thermal_photons}). Different hydrodynamical calculations of thermal photon 
production~\cite{dde_peressou,rasanen_sps_rhic,srivastava_sps_rhic,turbide,alam05}
%with initial entropy densities  fixed so as to reproduce the bulk hadron multiplicities at RHIC, %
can reproduce the PHENIX data assuming the formation of a radiating QGP with temperatures around 
$2\,\Tcrit$ ($T_0$ = 400 -- 600 MeV, corresponding to average temperatures 
$\mean{T_0}\approx$ 350 MeV over the source profile) at times $\tau_0$ = 2$R_A$/$\gamma\approx$ 0.2 fm/$c$  
right after the two colliding nuclei pass-through each other. A word of caution should be noted in that
the baseline ($T_{AA}$-scaled) prompt $\gamma$ spectrum at $p_T$ = 1 -- 4 GeV/$c$ is obtained 
from NLO calculations, because no measurement is yet available in this $p_T$ range. 
The confirmation of the existence of a thermal enhancement over the prompt component, will require 
a direct measurement of the $pp$ photon spectrum down to $p_T =$ 1 GeV/$c$.\\

\noindent
After subtracting the prompt $\gamma$ component from the inclusive direct photon spectrum, 
the local inverse slope parameter $\Teff$ in the range $p_T\approx$ 2 -- 4 GeV/$c$ 
can be used as a relatively good surrogate of the initial medium temperature~\cite{dde_peressou}. 
The combination of $\Teff$ with another global observable directly related to the entropy density $s$ 
would therefore allow one to determine the effective number of degrees of freedom $g$ of the 
system via the Stefan-Boltzmann ratio $g\propto s/T^{3}$. %for an ultrarelativistic gas of massless particles. 
Since the final charged particle rapidity density $\dNdeta$ is directly correlated with the initial entropy density of the system, 
by empirically studying the evolution of  $\geff\propto\dNdeta/\Teff^3$ versus $\Teff$ in different $AA$ 
centralities one can effectively study the evolution of the number of degrees of freedom and look for any threshold 
behaviour related to the sudden increase at the transition temperature and/or a flattening of $\geff$ for 
temperatures above $\Tcrit$. Such an approach has been tested in the context of a 2D+1 hydrodynamical 
model which effectively reproduces the hadron and photon data at RHIC~\cite{dde_peressou}. We found that 
one can clearly distinguish between the equation of state of a weakly interacting QGP and that of a system 
with hadron-resonance-gas-like EoS (i.e. with rapidly rising number of mass states with $T$). 
%More quantitative conclusions on the exact shape of the underlying EoS and/or 
%the absolute number of degrees of freedom of the produced medium require more detailed 
%theoretical studies as well as high precision photon data in $AuAu$ in the $p_T$ range where
%the QGP is expected to shine as well as baseline $pp$, $dAu$ collisions in the same range.
However, direct evidence of the parton-hadron phase change itself as a jump in $\geff$ around
$\Teff\sim\Tcrit$ %via the study of the centrality dependence of the hadron multiplicities and thermal photon slopes would 
can only be potentially visible in $AuAu$ reactions at {\it lower} center-of-mass energies 
($\sqrt{s_{NN}}\approx$ 20 -- 65 GeV)~\cite{dde_peressou2}. At the LHC, the expected 
temperatures $\mathscr{O}$(1 GeV) reached in central $PbPb$ will also produce a significant 
thermal photon signal up to $p_T\approx$ 6 GeV/$c$~\cite{yr_lhc_photons} which can be used to 
determine the thermodynamical conditions in the {\it plateau} regime of the $s/T^3$ EoS, 
closer to the ideal-gas limit than at RHIC.

%%%%%%%%%%%%%%%%%%%%%%%%%%%%%%%%%%%%%%%%%%%%%%%%%%%%%%%

\section{Critical temperature and energy density: Anomalous $\jpsi$ suppression}
\label{sec:jpsi}

\begin{figure}[htb]
\centering
\includegraphics[width=10cm,height=7cm]{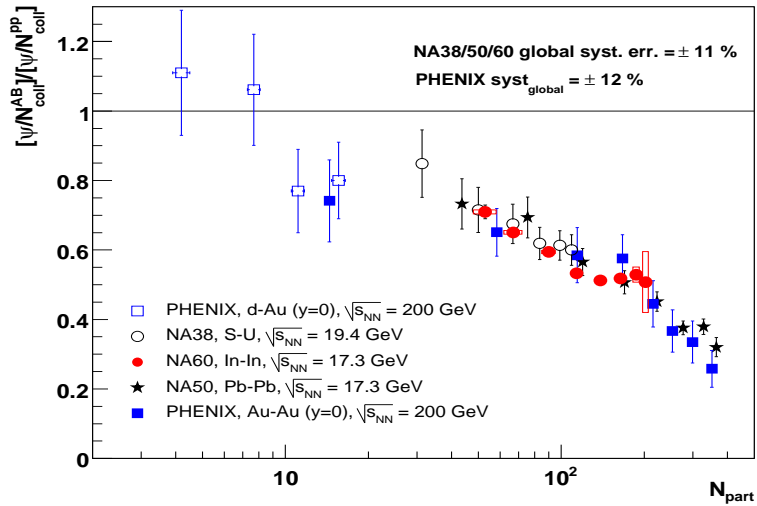}\hskip -0.1cm
\vskip -0.4cm
\caption{$\jpsi$ nuclear modification factor versus centrality~\cite{clourenco_hp06} (given 
by the number of participant nucleons in the collision) measured in $AA$ and $p(d)A$ collisions 
at SPS~\cite{na38_jpsi,na50_jpsi,na60_jpsi} and RHIC~\cite{phnx_dAu_jpsi,phnx_jpsi}.}
%$\jpsi$ survival probability~\protect\cite{satz05} (Eq.~(\ref{eq:R_dA})) versus energy density
%in different $AA$ reactions  ... ($\varepsilon$ is given by the Bjorken estimate, Eq.~(\ref{eq:Ebj}), 
%with $\tau_0$ = 1 fm/$c$ an (energy-independent) $\jpsi$ formation time).}
\label{fig:RAA_jpsi}
\end{figure}

The study of heavy-quark bound states in high-energy $AA$ collisions has been long since proposed  
as a sensitive probe of the thermodynamical properties of the produced medium~\cite{matsui_satz}. 
Analysis of quarkonia correlators and potentials in finite-$T$ lattice QCD indicate that the different 
charmonium and bottomonium states dissociate at temperatures for which the colour (Debye) screening radius 
of the medium falls below their corresponding $Q\bar{Q}$ binding radius. Recent lattice analyses of the 
quarkonia spectral functions~\cite{QQbar_latt} indicate that the ground states ($\jpsi$ and $\ups$) survive 
at least up to $T\approx 2\Tcrit$ whereas the less bounded $\chi_c$ and $\psi'$ melt near $\Tcrit$. 
Experimental confirmation of such a threshold-like dissociation pattern would provide a %an experimentally testable 
direct means to determine the transition temperature reached in the system and their comparison 
to {\it ab initio} lattice QCD predictions. A significant amount of experimental data on $\jpsi$ production 
in different $p(d)A$ and $AA$ collisions has been collected at SPS~\cite{na38_jpsi,na50_jpsi,na60_jpsi} and 
RHIC~\cite{phnx_jpsi}. The corresponding nuclear modification factors compiled in~\cite{clourenco_hp06}
are shown in Fig.~\ref{fig:RAA_jpsi} as a function of $N_{part}$. 
%The so-called ``survival probability'' version of  Eq.~(\ref{eq:R_dA}), points out to an ``anomalous suppression'' 
%of total $\jpsi$ yields beyond the ``normal'' suppression in cold nuclear matter for increasing energy densities  $\varepsilon$ 
%(Fig.~\ref{fig:RAA_jpsi}, left).
%Since $T\approx (\varepsilon/6)^{3/4}$, the highest temperatures attained at RHIC are of the order 
%of $T\approx 2\Tcrit$ and, thus, close to the threshold for direct $\jpsi$ suppression in the plasma.
The surprisingly similar amount of $\jpsi$ suppression observed at SPS and RHIC energies (with expected 
temperature differences of a factor of $\sim$2) has been interpreted in a sequential-dissociation 
scenario~\cite{satz05} where the $\jpsi$ survives up to $T\approx 2\Tcrit$ in agreement with the lattice predictions, 
and the observed suppression at both c.m. energies is just due to the absence of (30\% and 10\%) feed-down 
decay contributions from $\chi(1P)$ and $\psi'(2S)$ resonances which melt already at $T\approx\Tcrit$. 
The confirmation of such an interpretation would set an upper limit of $T\lesssim 2\Tcrit\approx$  400 MeV
for the temperatures reached at RHIC.

\noindent
Other explanations of the comparatively low depletion of $\jpsi$ yields at RHIC have been 
put forward based on a much stronger direct $\jpsi$ suppression (at temperatures close to $\Tcrit$) 
combined with $c\bar{c}$ pairs regeneration from the abundant charm quarks\footnote{10 charm pairs 
are produced on average in a central $AuAu$ collision at the top RHIC energy.} in the dense medium~\cite{qqbar_reco}.
%a better control on the ``normal'' suppression in cold nuclear matter and without a
%precise determination of the $\jpsi$ feed-down rates from $\chi(1P)$ and $\psi'(2S)$.
%direct comparison: experimental results vs. quantitative QCD predictions
The LHC measurements will be crucial to resolve this issue. A strongly suppressed $\jpsi$
yield in $PbPb$ at 5.5 TeV -- where the expected initial temperatures will be well above 2$\Tcrit$ -- 
would support the sequential-screening scenario, whereas recombination models predict a strong 
enhancement due to the larger density of $c\bar{c}$ pairs in the medium. In addition, the abundant 
production of the $\ups(1s,2s,3s)$ states at LHC energies will open up a unique opportunity
to study the threshold dissociation behaviour of the whole bottomonium family. The $\ups$ is expected
to survive up to $4\,\Tcrit$ and, therefore, {\it direct} suppression of the $b\bar{b}$ 
ground-state would be indicative of medium temperatures around 1 GeV at the LHC.

%%%%%%%%%%%%%%%%%%%%%%%%%%%%%%%%%%%%%%%%%%%%%%%%%%%%%%%

\section{Chiral symmetry restoration: In-medium vector mesons}
\label{sec:chiral}

In-medium modifications of the spectral function (mass, width) of the light vector mesons ($\rho$, 
$\omega$, and $\phi$) have been proposed as a promising signature of the (approximate) 
restoration of  chiral symmetry in the $u,d,s$ quark sector~\cite{chiral_symm,brown_rho}. 
In the QCD vacuum %leads to the dressing of the $u,d,s$ quarks with a constituent mass which 
the spontaneous breaking of chiral symmetry manifests itself in the hadron spectrum 
through the mass-splitting of ``chiral partners'' i.e. between states of opposite parity but 
equal quantum numbers. Chiral symmetry breaking leads, in the mesonic sector, to the non-degeneracy 
of the pseudo-/scalar ($\pi-\sigma$) and axial-/vector ($a_1-\rho$) channels. For temperatures above the 
chiral transition, massless left- and right-handed quarks will decouple and one expects to observe a 
gradual disappearance of the mass-splitting, leading to a shift of the masses of vector mesons and their chiral partners.
% degeneracy of the spectral densities within a chiral multiplet. %are expected to become degenerate. 
The $\rho$ meson is an excellent candidate for the experimental study of in-medium spectral functions in 
$AA$ collisions due to (i) its short lifetime ($\tau_0$ = 1.3 fm/$c$) allowing it to decay before regaining 
its vacuum spectral shape, and (ii) its (rare but detectable) dilepton decay branching ratio 
($\Gamma_{l+l-}\approx 5 \cdot 10^{-5}$) which is unaffected by final-state interactions with the 
surrounding environment.\\

\begin{figure}[htb]
\centering
\includegraphics[width=10.cm,height=7.cm]{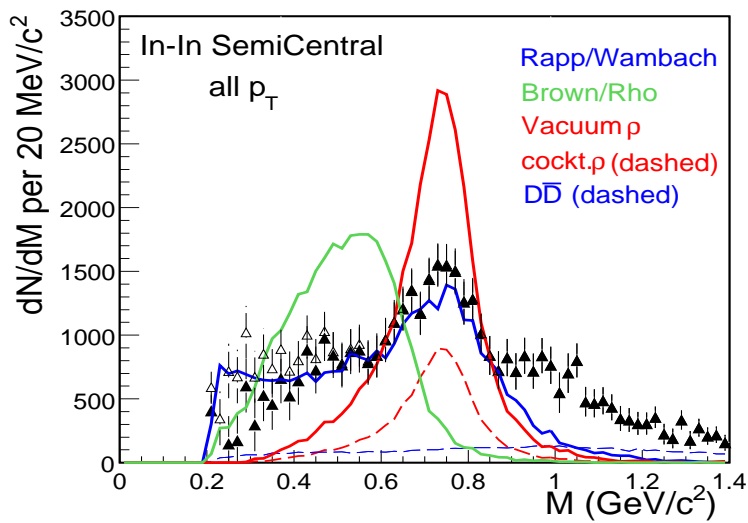}
\vskip -0.cm
\caption{``Excess'' invariant $\mu^+\mu^-$ mass spectrum in semi-central $InIn$ collisions at 
$\sqrtsnn$~=~17.3 GeV~\cite{na60_rho} compared to the expected $\rho$ line-shape in different 
theoretical scenarios: vacuum $\rho$, collisional broadening~\cite{chiral_symm,hvh_rapp_rho}, 
and mass drop~\cite{brown_rho}.}
\label{fig:rho_mass}
\end{figure}

\noindent
The NA60 experiment has recently studied a high-statistics sample of low mass muon pairs in $InIn$ collisions
at $\sqrtsnn$ = 17.3 GeV~\cite{na60_rho}. The measured $\rho$ spectrum, after subtraction of all other sources
of opposite-sign muon pairs, is peaked at the nominal (free) mass ($m_{\rho}$ = 0.77 GeV/$c^2$) but its
width is substantially broadened for increasing centralities (Fig.~\ref{fig:rho_mass}).
The $\rho$ spectral shape is better reproduced by models which consider in-medium width broadening 
due to interactions in a hot and dense {\it hadronic} medium close to the expected phase boundary 
($T\sim$ 190 MeV)~\cite{hvh_rapp_rho}, than by a long-standing prediction based on a downwards mass-shift 
coupled directly to the melting of the chiral condensate~\cite{brown_rho}. Such a result sets new constraints 
on the possible realization of chiral symmetry in QCD matter. At RHIC energies, the factor of two larger temperatures 
reached compared to SPS prefigure that the $\rho$ should %the reached temperatures are a factor $\sim$2 larger than at SPS, 
exhibit a completely ``melted'' line-shape if, as expected from the lattice, the chiral and deconfinement transitions 
occur at the same $\Tcrit$. The $\rho$ measurement in the dielectron channel will be possible in PHENIX
with the recently installed hadron-blind-detector~\cite{HBD} which will allow to suppress by two orders 
of magnitude the large combinatorial background arising from light-meson Dalitz decays and photon 
conversions, while preserving 50\% of the signal.

%%%%%%%%%%%%%%%%%%%%%%%%%%%%%%%%%%%%%%%%%%%%%%%%%%%%%%
%\clearpage
\section*{Summary}
\label{sec:summary}
High-energy  collisions of heavy ions provide the only existing method today to explore empirically 
the phase diagram of QCD at extreme values of temperature, density and low-$x$. We have reviewed 
the experimental and phenomenological progress in nucleus-nucleus collisions at BNL-RHIC 
($\sqrtsnn\approx$ 20 -- 200 GeV) and CERN-SPS ($\sqrtsnn\approx$ 20 GeV) in recent years 
with a particular emphasis on the modifications suffered by different hard probes in the QCD medium 
that is produced, compared to baseline measurements in proton-proton or proton(deuteron)-nucleus collisions. 
The observed modifications allow us to obtain direct information on fundamental (thermo)dynamical 
properties of strongly interacting matter.\\

\noindent 
The reduced $AuAu$ hadron multiplicities and the depleted yields of semihard hadrons 
($p_T\approx$ 1 -- 4 GeV/$c$) at forward rapidities in $dAu$ collisions indicate that the initial conditions 
of the nuclei accelerated at RHIC energies are consistent with Colour-Glass-Condensate approaches that
model the hadronic parton distributions as a low-$x$ saturated gluon wavefunction evolving according 
to non-linear QCD equations. These new data -- complementary to the rich HERA results on the partonic 
structure of the proton -- shed new light on the high-energy limit of QCD, a physics topic not only appealing 
in its own right but an essential ingredient for any serious attempt to compute a large variety of hadron-, 
photon- and neutrino- scattering cross sections at increasingly large energies.\\
%, with implications for example in ultra-high energy cosmic ray physics.\\

\noindent 
The robust radial and elliptic flows seen for all identified hadron species up to $p_T\approx$ 2 GeV/$c$ 
in $AuAu$ reactions at $\sqrtsnn$ = 200 GeV are remarkably well described by {\it ideal} relativistic 
hydrodynamics calculations which model the expanding system starting with a realistic QGP equation-of-state 
with initial energy densities $\varepsilon_0\approx$ 30 GeV/fm$^3$ at thermalization times $\tau_0\approx$ 0.6 fm/$c$.
%all hadron species (including the heavier phi, xi, omega) boosted with a common velocity, strongly correlated 
% with the maximum pressure direction. strong v2 dependence on the hadron mass nicely matches with the 
%predictions of fluid dynamical simulations of the expansion of an ellipsoidal fireball with realistic QGP equation-of-state.
%the calculations indicate that the elliptic flow builds up due to pressure gradients generated
%in multiple scattering in the equilibration process before 1 fm/c when the
%energy of the matter exceeds 1 GeV/fm3. the calculations also show sensitivity to the
%equation of state but a direct extraction of $\varepsilon(P)$ is not straightforward.
Detailed hydro-data comparisons for various differential observables -- in particular the elliptic flow 
parameter $v_2(p_T)$ for different light and heavy hadron species -- indicate that the system exhibits 
very small viscosities (i.e. very short mean free paths) and is strongly coupled at variance 
with the anticipated QGP paradigm of a weakly interacting gas of relativistic partons.
In the intermediate $p_T\approx$ 2 -- 4 GeV/$c$ range, baryons have enhanced yields and 
flows compared to mesons pointing to a novel channel for hadronization based on 
constituent-quark coalescence in a dense partonic medium. The existence of a new mechanism 
of hadron formation in heavy-ion reactions at transverse momenta of a few GeV/$c$, apart 
from standard parton fragmentation, offers new ways to probe the space-time dynamics 
of confinement in different QCD environments.\\

\noindent 
In the high-$p_T$ sector, leading hadrons (but not colour-blind prompt $\gamma$) are suppressed 
by up to a factor of $\sim$5 in the range $p_T\approx$ 4 -- 20 GeV/$c$ compared to 
perturbatively-scaled proton-proton spectra. This result is in agreement with non-Abelian energy loss models 
that assume that the produced partons traverse a medium with very large parton rapidity densities 
$dN^g/dy\approx$ 1100 and transport coefficients $\qhat\approx$ 10 GeV$^2$/fm. The energy lost 
by the quenched parton in the medium apparently shows up in a very unconventional conical-like azimuthal 
profile of secondary hadrons ($p_T\approx$ 1 -- 3 GeV/$c$) in the away-side hemisphere of high-$p_T$ trigger 
hadrons. Interpretations of this preferential azimuthal emission at $\Delta\phi\approx\pi\pm$ 1.1, as 
caused by the generation of a Mach-cone boom by a supersonic parton propagating through the dense 
system, yield average speeds of sound $\mean{c_s} = \cos(\theta_{M})\approx$ 0.45 not far from those 
expected from lattice QCD for deconfined quark-gluon matter.\\

\noindent 
In the electromagnetic sector, preliminary real and virtual $\gamma$ spectra in central $AuAu$ in 
the range $p_T$ = 1 -- 14 GeV/$c$  can be described as the sum of a perturbative (prompt) 
photon contribution plus a secondary component of thermal origin. Hydrodynamics calculations 
can reproduce the data assuming the formation of a radiating QGP with average temperatures 
$2\,\Tcrit\approx$ 350 MeV. Such temperature values seem to be consistent with the observation 
that the amount of $\jpsi$ suppression at RHIC and SPS is about the same as expected from lattice-based 
calculations that predict a survival of the $c\bar{c}$ ground state up to $T\approx 2\Tcrit$ but a ``melting'' 
of the $\chi_c$ and $\psi'$  (which feed-down at a $\sim$40\% level to the $\jpsi$) near $\Tcrit$. At the CERN
SPS, recent NA60 results on the $\rho$ spectral function in central $InIn$ collisions at $\sqrtsnn$ = 17.3 GeV 
indicate that the width of the vector meson is substantially broadened in the medium. Theoretical calculations
indicate that most of the broadening can be accounted for by collisional effects in a hot and dense 
hadronic medium with initial temperatures close to $\Tcrit\approx$175 MeV (note that such a result is 
also consistent with the observed hadron abundances at SPS which indicate that the system reaches 
chemical equilibrium at temperatures around 160 MeV).\\
%The above topics  are deeply related to the unsolved problems of the strong interaction: confinement
%dynamics, ... thus they offer an opportunity to study the dynamics of ... and new ways to probe 
%confinement dynamics in space and time.

\noindent 
The overall scenario taking form from the wealth of recent experimental data suggests, on the one hand, 
that heavy-ion collisions at RHIC energies produce a strongly interacting liquid-like QGP with very large
initial parton rapidity densities $dN^g/dy\approx$ 1100, temperatures $2\,\Tcrit\approx$ 350 MeV and 
very low shear viscosities. On the other hand, systems produced at SPS seem to be only partially thermalised 
-- according to the lower measured collective anisotropic flow compared to RHIC --, have initial
$dN^g/dy\approx$ 400 %(to account for a modest high $p_T$ meson suppression)
and temperatures around the phase boundary at $\Tcrit\approx$ 175 MeV.  However, it is fair to 
acknowledge that among the existing signals there is yet no incontrovertible ``textbook'' figure {\it proving} 
the formation of a thermalised extended medium consisting of deconfined and chirally-symmetric quarks 
and gluons. One such evidence would be a direct empirical observation of a jump in the number 
of effective degrees of freedom at the phase change as expected by EoS calculations in the lattice. 
Since RHIC top energies seem to produce systems at twice $\Tcrit$ and SPS data point to conditions just at 
the predicted phase change, the expected jump is likely to be observed  (e.g. via precise studies 
of the correlation of thermal photon slopes with the global hadron multiplicities) in a next phase of RHIC
running at intermediate energies ($\sqrtsnn\approx$ 20 -- 62 GeV) and high luminosities. Likewise, 
confirmation of the concurrent chiral and deconfinement transitions in QCD matter will require e.g. precise
measurements of the $\rho$ spectral shape (and, ideally, that of its chiral partner $a_1$) at RHIC energies. 
Lower-energy runs at RHIC, as well as at the projected CBM facility~\cite{cbm}, will access also the region
of large baryon densities around the QCD critical point. Finding signs of the tri-critical point at relatively 
high temperatures would indicate that the smooth cross-over changes to a first order phase transition 
at higher baryon densities, a result of relevance for the conditions prevailing in the core of neutron 
(and other compact) stars. Direct validation of the strongly-coupled interpretation of the medium formed 
at RHIC and potential observation of the anticipated {\it weakly} interacting quark-gluon plasma require 
key measurements in $PbPb$ at 5.5 TeV at CERN-LHC where the initial temperatures $\mathscr{O}$(1 GeV) 
should be large enough to observe the direct melting of ground-state quarkonia resonances. In addition, 
at the LHC, the longer duration of the QGP phase and the much abundant production of other hard probes (in particular 
parton energy loss results for {\it fully} reconstructed, $\gamma$- or $Z$-tagged, and flavour-identified jets) 
thermal photons, $v_2$ flow parameter, etc. will likely result in indisputable probes of the deconfined medium 
much less dependent on details of the later hadronic phase.\\

\noindent 
The experimental advances in the last years have been paralleled by significant progresses in
the theoretical description of high-density QCD matter. %Theoretical methods to study the full richness of the theory: 
Lattice methods are increasingly more refined and powerful to describe not only the infrared 
collective dynamics (EoS, critical parameters) but also the in-medium correlators (quarkonia). 
Effective field theories have been developed in specific domains such as the Colour-Glass-Condensate
which effectively describes the high-energy (low-$x$) limit of QCD. Perturbative calculations have 
substantially improved the description of the interaction of hard probes with hot and dense 
quark-gluon matter. Last but not least,  duality approaches based on the application of AdS/CFT 
correspondence between weakly coupled gravity and strongly coupled QCD-like systems, are 
providing new powerful insights on dynamical properties that cannot be directly treated by either 
perturbation theory or lattice methods while simultaneously opening novel directions for 
phenomenological studies and experimental searches.\\

\noindent
The impressive experimental and theoretical advances triggered by the wealth of high-statistics,
high-quality data collected in ultrarelativistic nucleus-nucleus collisions at RHIC and SPS, have 
significantly expanded the knowledge of many-body QCD at extreme conditions of temperature, 
density and low-$x$. Those studies -- which will be substantially extended in the upcoming LHC
(and likely RHIC-II) nucleus-nucleus and proton-nucleus programme -- go beyond the strict realm 
of the strong interaction and shed light on a vast ramification of fundamental physics problems. 
Knowledge of the collective behaviour of many-parton systems is of primary importance not only 
to address basic aspects of the strong interaction such as the nature of confinement or the mechanism 
of mass generation via chiral symmetry breaking, but to ascertain the high-energy limit of all scattering 
cross sections involving hadronic objects, the inner structure of compact stellar objects, or the evolution 
of the early universe between the electroweak transition and primordial nucleosynthesis.

%%%%%%%%%%%%%%%%%%%%%%%%%%%%%%%%%%%%%%%%%%%%%%%%%%%%%%

\section*{Acknowledgments}

\noindent 
Special thanks due to F.~Antinori, N.~Armesto, W.~Busza, A.~de~Roeck, D.~Denegri, J.~Harris,
B.~Jacak, P.~Jacobs,  C.~Louren\c{c}o, G.~Rolandi, C.~Salgado, J.~Schukraft, Y.~Schutz, W.~Wyslouch, 
and B.~Zajc for a careful reading of the manuscript, informative discussions and useful suggestions.
%I gladly acknowledge informative discussions with ...
% antinori, armesto, busza, gyulassy, harris, jacak, lourenco, mclerran, mjt, roland, salgado, schukraft, schutz, wyslouch, wiedemann, zajc, roy, jacobs
% de roeck, sphicas, janot, denegri, rolandi
% for comments on the text.
This work is supported by the 6th EU Framework Programme contract MEIF-CT-2005-025073.

%%%%%%%%%%%%%%%%%%%%%%%%%%%%%%%%%%%%%%%%%%%%%%%%%%%%%%

%%%%%%%%%%%%%%%%%%%%%%%%%%%%%%%%%%%%%%%%%%%%%%

\end{document}